\documentclass{emulateapj}
\newcommand{\bvec}[1]{\textbf{#1}}

\newcommand{\symvec}[1]{\mbox{\boldmath${#1}$}}

\shorttitle{LSST Weak Lensing Simulation}
\shortauthors{Jee et al.}

\begin{document}

\title{TOWARDS PRECISION LSST WEAK-LENSING MEASUREMENT - I: IMPACTS OF \\ ATMOSPHERIC TURBULENCE AND OPTICAL ABERRATION}

\author{M. JAMES JEE\altaffilmark{1} AND J. ANTHONY TYSON\altaffilmark{1}}

\begin{abstract}
The weak-lensing science of the Large Synoptic Survey Telescope (LSST) project drives the need to carefully model and separate the instrumental artifacts from the intrinsic shear signal caused by gravitational lensing. The dominant source of the systematics for all ground-based telescopes is the spatial correlation of the point spread function (PSF) modulated by both atmospheric turbulence and optical aberrations in the telescope and camera system.
In this paper, we present a full field-of-view simulation of the LSST images by modeling both the atmosphere and the system optics with the most current
data for the telescope and camera specifications and the environment. To simulate the effects of atmospheric turbulence, we generated six-layer Kolmogorov/von K{\'a}rm{\'a}n phase screens with the parameters estimated from the on-site measurements.
LSST will continuously sample the wavefront, correcting the optics alignment and focus.
For the optics, we combined the ray-tracing tool ZEMAX and our simulated focal plane data to introduce realistic residual aberrations and focal plane height variations. 
Although this expected focal plane flatness deviation for LSST is small compared with that of other existing cameras, the fast f-ratio of the LSST optics makes this focal plane flatness variation and the resulting PSF discontinuities across the CCD boundaries significant challenges in our removal of the PSF-induced systematics. We resolve this complication by performing principal-component-analysis (PCA) CCD-by-CCD, and interpolating the basis functions derived from the analysis 
using conventional polynomials.
We demonstrate that this PSF correction scheme reduces the residual PSF ellipticity correlation below $10^{-7}$ over the cosmologically interesting (dark matter dominated) scale $10\arcmin-3 \degr$. From a null test using Hubble Space Telescope (HST) Ultra Deep Field (UDF) galaxy images without input shear, we verify that the amplitude of the galaxy ellipticity correlation function, after the PSF correction, is consistent with the shot noise set by the finite number of objects.
We conclude that the current optical design and specification for the accuracy in the focal plane assembly are sufficient to enable the
control of the PSF systematics required for weak-lensing science with LSST.
\end{abstract}

\altaffiltext{1}{Department of Physics, University of California, Davis, One Shields Avenue, Davis, CA 95616}


\clearpage

\section{INTRODUCTION \label{section_introduction}}

The Large Synoptic Survey Telescope (LSST) has been designed to provide a deep six-band (0.3-1.1 $\micron$) astronomical
imaging survey of more than $20,000$ square degrees of the southern sky.  Using active optics, the 8.4-meter aperture and 9.6 square degree
field-of-view of the telescope will allow $\sim1000$ visits to each patch of sky in ten years with a final
depth reaching $r\sim27.5$ and a mean delivered PSF better than 0.7$\arcsec$ (LSST Science Collaborations et al. 2009). 
The LSST project will deliver fully-calibrated, science quality images,
catalogs, and derived data products to the US public with no proprietary period. 
We refer readers to LSST Science Book II (LSST Science Collaboration 2009) for extensive discussions on the science that the survey
will enable.

Among the most critical science applications that the LSST will revolutionize is weak gravitational lensing. 
The key observable in weak gravitational lensing is the shape of distant galaxies weakly distorted by foreground masses. Because each galaxy is sheared only by
a small amount, the lensing signal must be extracted from an average over a population of galaxies. Consequently, the total number of resolved galaxies at the
usable surface brightness limit
is one 
useful measure to quantify the effectiveness of an astronomical survey for weak-lensing science. Another important factor is the number density of
usable galaxies (shear noise weighted effective number), which determines the lower limit of the angular scale that lensing can probe. The LSST project is designed to maximize
the scientific return balancing both the total number (area) and the number density within the ten-year nominal mission period.
A sky coverage of apptroximately $20,000$ square degrees is required to keep the sample variance below the statistical limit, and the
high-resolution ($<0.7\arcsec$ median delivered seeing), deep imaging ($r\sim27.5$), providing $>40$ high S/N galaxies per square arcmin enabling a reliable
probe of matter structure on galaxy cluster scales. Residual uncorrected PSF shape variation requirements are summarized in the Science Requirement
Documents: http://www.lsst.org/filesdocs/SRD.pdf.

These performance requirements determine the desired level of accuracy in point spread function (PSF) correction. The effect of the PSF on
the galaxy shape measurement cannot be overemphasized. Anisotropic PSFs mimic gravitational lensing signal, generating a coherent alignment
of galaxy shapes. Even without this anisotropy, PSFs circularize shapes of barely resolved galaxies
and dilute the intrinsic lensing signal. For various reasons, the observed
PSF changes in size and anisotropy across the field and also over time. Moreover, we need the PSF extrapolated to the positions of
galaxies, whereas the PSF model is constructed from the high S/N stars that are much less densely distributed.
Therefore, in addition to very low surface brightness limits, a key factor in precision lensing lies in one's ability to carefully
model and remove the complicated effects of PSFs.

Although the seeing at the LSST site is among the best in existing ground-based facilities, the accurate description of the PSF
is highly challenging because of the optical and focal plane design. As described below, the telescope optics and camera are continuously aligned via 
wavefront sensing.  Optimized for the unprecedented large field of view, the effective
$f$-ratio of LSST is small $f/\sim1.2$, which makes the optical aberration highly sensitive to alignment and focal plane deviation. Because
LSST's focal plane will be tiled with 189 4k $\times$ 4k CCDs, any height fluctuation both across and within the CCDs will translate into a very
complicated PSF pattern, which is characterized by abrupt changes across the CCD gaps and smooth, but possibly high-frequency variations within the CCDs.
Any sub-optimal modeling of these PSFs will leave systematic residuals on the scales of the CCD sizes and the ``potato chip'' effect\footnote{
The substrate bonding and packaging process of the CCD fabrication induces a height variation, which leads to non-negligible focus variation (thus 
variation in aberration) within a CCD.}.
These
systematic residuals, mimicking lensing signals, will, of course, hamper the correct interpretation of the lensing analysis.

Unfortunately, with the existing algorithms the ability to precisely describe the PSF alone does not guarantee the success in the extraction
of gravitational shear to the accuracy that future LSST-like weak-lensing surveys require. This obstacle is well noted in the recent
large collaborative shear measurement campaign GREAT08 Challenge (Bridle et al. 2010). The campaign let the participants perform
blind shear measurements on simulated images, where the PSF was known, but the input shear was unknown. Although many algorithms
were shown to provide the accuracy for high S/N images
suitable for the existing survey data, no method came close to the target accuracy $Q\sim1000$ when the noise level of the images matched
the realistic value (see Bridle et al. 2010 for the definition of $Q$). Nevertheless, it is important to note that the main challenge is
purely mathematical/statistical and thus improvable as more mathematical sophistication is incorporated in the algorithms.
For example, Bernstein (2010) claims that when the bias arising 
when the true galaxy profiles do not match the models being fit is properly addressed, the resulting algorithm can achieve $Q\sim3000$
for high S/N galaxies of the GREAT08 sample. 

We launched the LSST shear calibration project in order to diagnose the key factors in the telescope and camera
engineering specifications affecting the weak-lensing science and to develop
new algorithms for optimal shear extraction in the presence of different combinations of systematics. The current paper is the first in the series of this topic with an emphasis on the realization, characterization, and reconstruction of LSST PSFs. 

The paper is organized as follows. In \textsection\ref{section_lsst_optics}, we provide a brief review on the telescope optics. 
Our implementation of the atmospheric turbulence is
described in \textsection\ref{section_atmosphere}. In Section \textsection\ref{section_focal_plane}, we discuss the focal plane design of LSST and the resulting behavior of the PSF. \textsection\ref{section_pca} presents our algorithm to measure and reconstruct the PSFs. In \textsection\ref{section_simulation}, we describe our simulation. Finally, the detailed comparison between the observed and modeled PSFs is presented in
\textsection\ref{section_analysis} before the conclusion in \textsection\ref{section_conclusion}.

\section{LSST ENVIRONMENT AND OPTICS}
\label{section_lsst_optics}
\begin{figure}
\includegraphics[width=9cm]{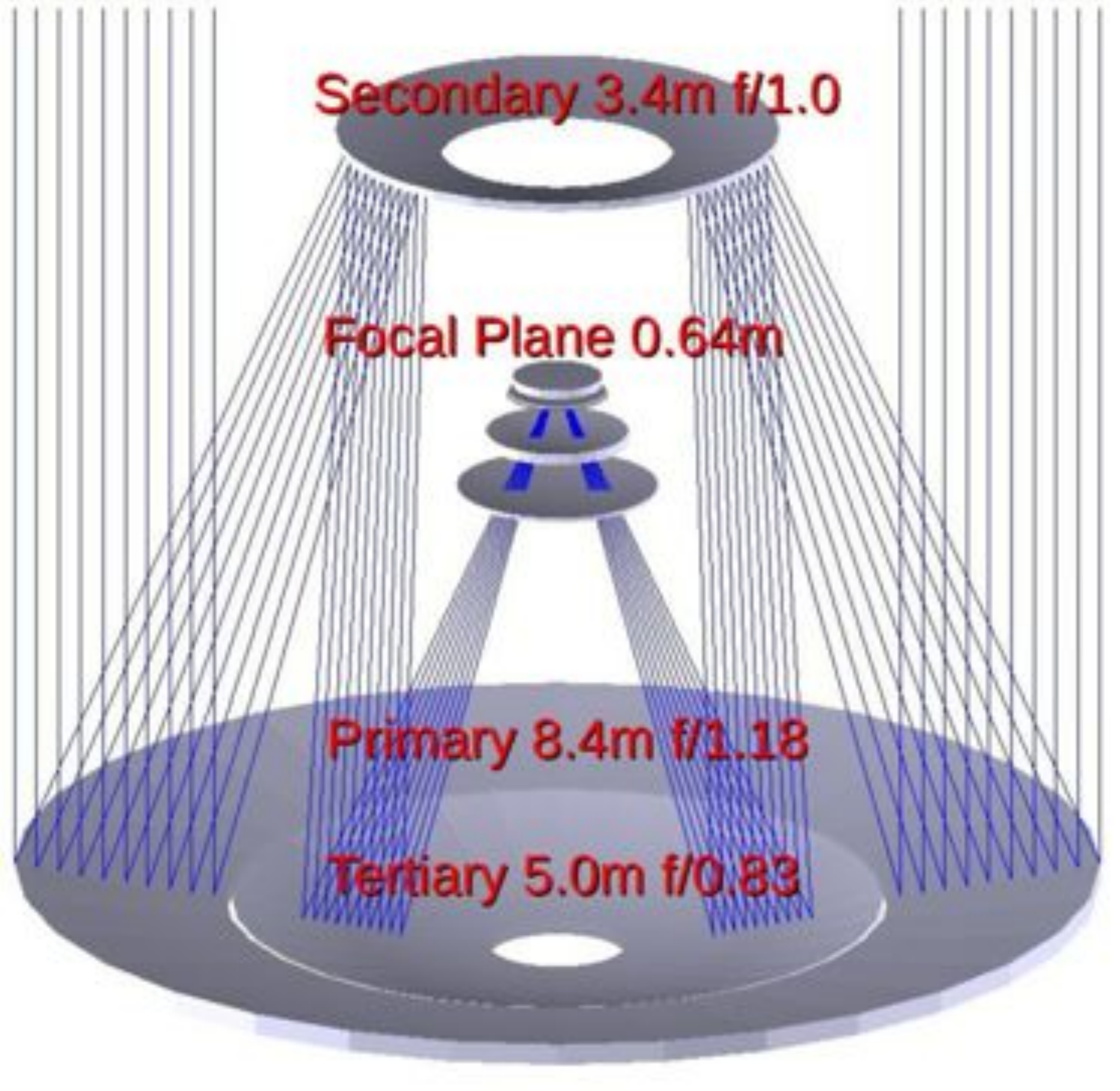}
\includegraphics[width=9cm]{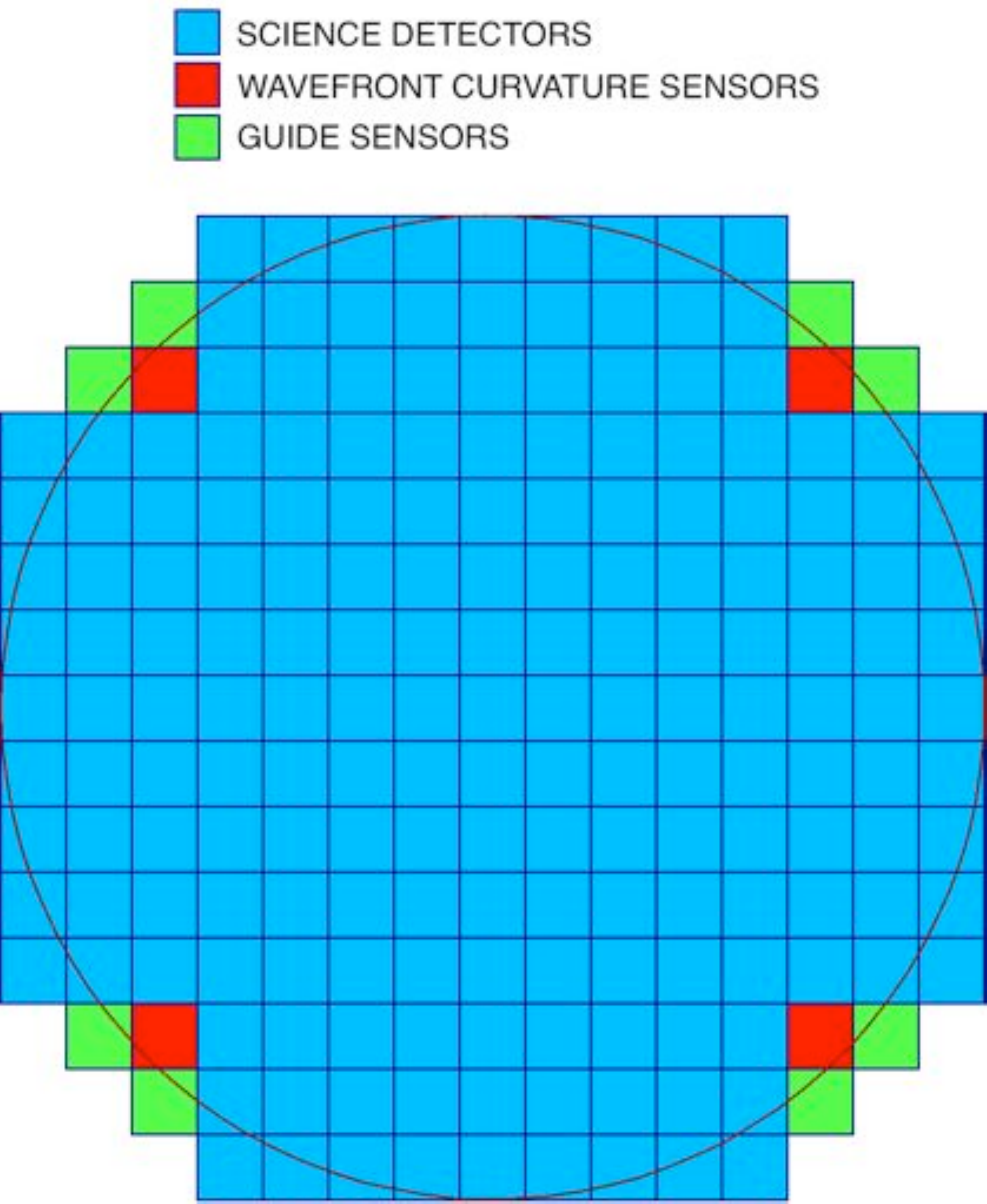}
\caption{LSST optical design. The telescope design (left) is a modified Paul-Baker three mirror system, which produces a uniform image quality
across a large field of view. The relative positions of the primary and the tertiary mirrors were adjusted during the
design process so that their surfaces meet with no axial discontinuity at a cusp, allowing the Primary and the Tertiary
to be fabricated from a single substrate. The 3.4-m convex secondary mirror
has a 1.8 m inner opening, through which the LSST camera is inserted at the focal plane. 
The focal plane (right) is tiled with 189 4k$\times$4k CCDs. There are four wavefront sensors and
eight guide sensors, located in the four corners of the CCD array.
All optics and the camera are corrected 
with active optics wavefront curvature sensing. The red solid circle represents the 3.5\degr~diameter field of view boundary.
\label{fig_lsst_optics}}
\end{figure}

Figure~\ref{fig_lsst_optics} shows the optical design of the LSST, which is a modified Paul-Baker three-mirror system. 
The active optics telescope consists of an 8.4 m f/1.18 concave primary, a 3.4 m f/1.0 convex secondary, and 5.2 m f/0.83 concave tertiary mirrors.
The final focal ratio f/1.23 (focal length of 10.3 m)
gives a plate scale $0.0197\arcsec/\micron$ over the 64 cm diameter focal plane ($3.5\degr \times 3.5\degr$).
The 27\% obscuration yields a total effective light collecting area of 35 $\mbox{m}^2$, which corresponds to an effective
6.7 m diameter clear circular aperture. Current wide field telescopes were not designed for the stringent PSF demands of LSST weak lensing goals.

\begin{figure}
\includegraphics[width=9cm]{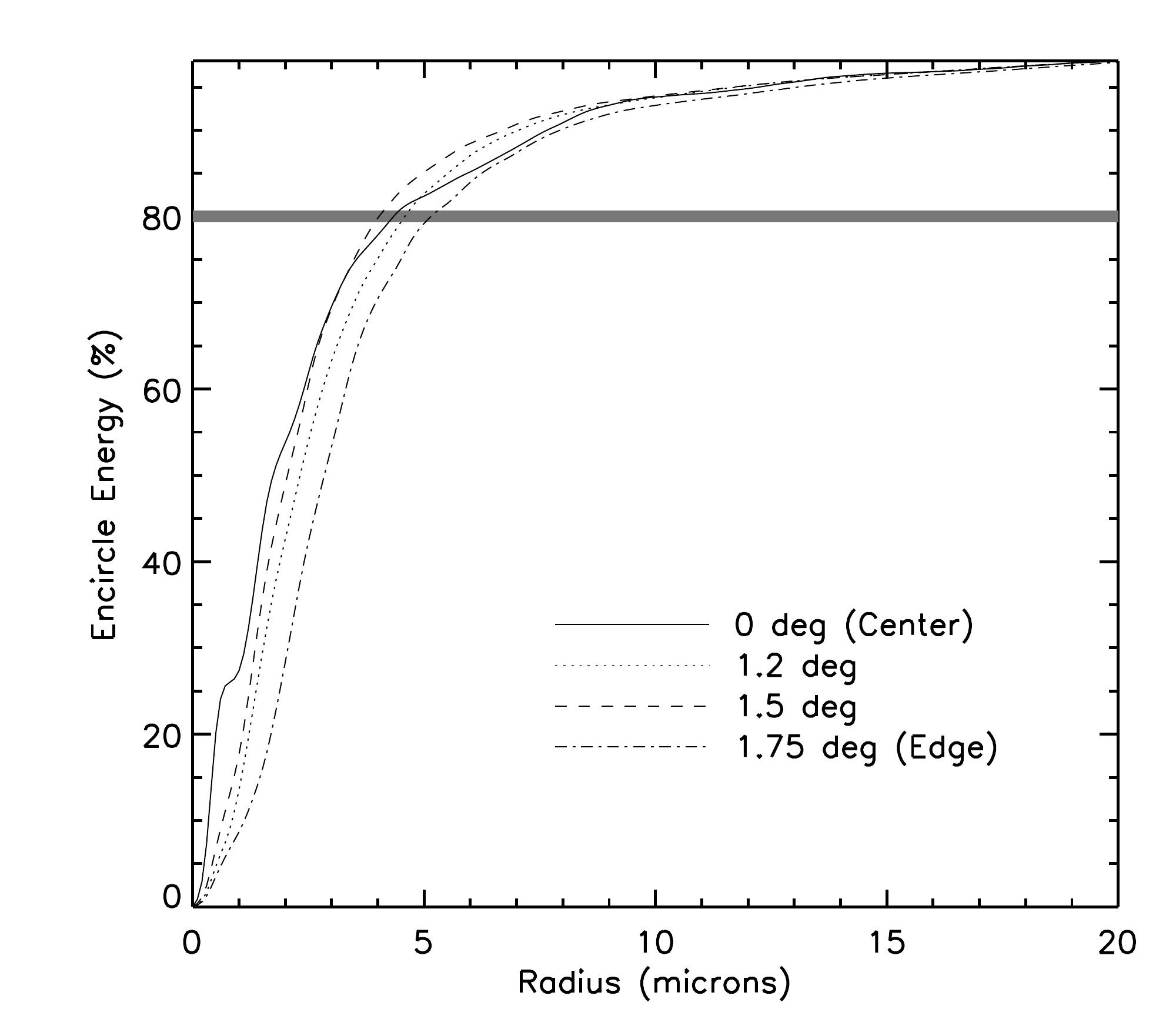}
\includegraphics[width=9cm]{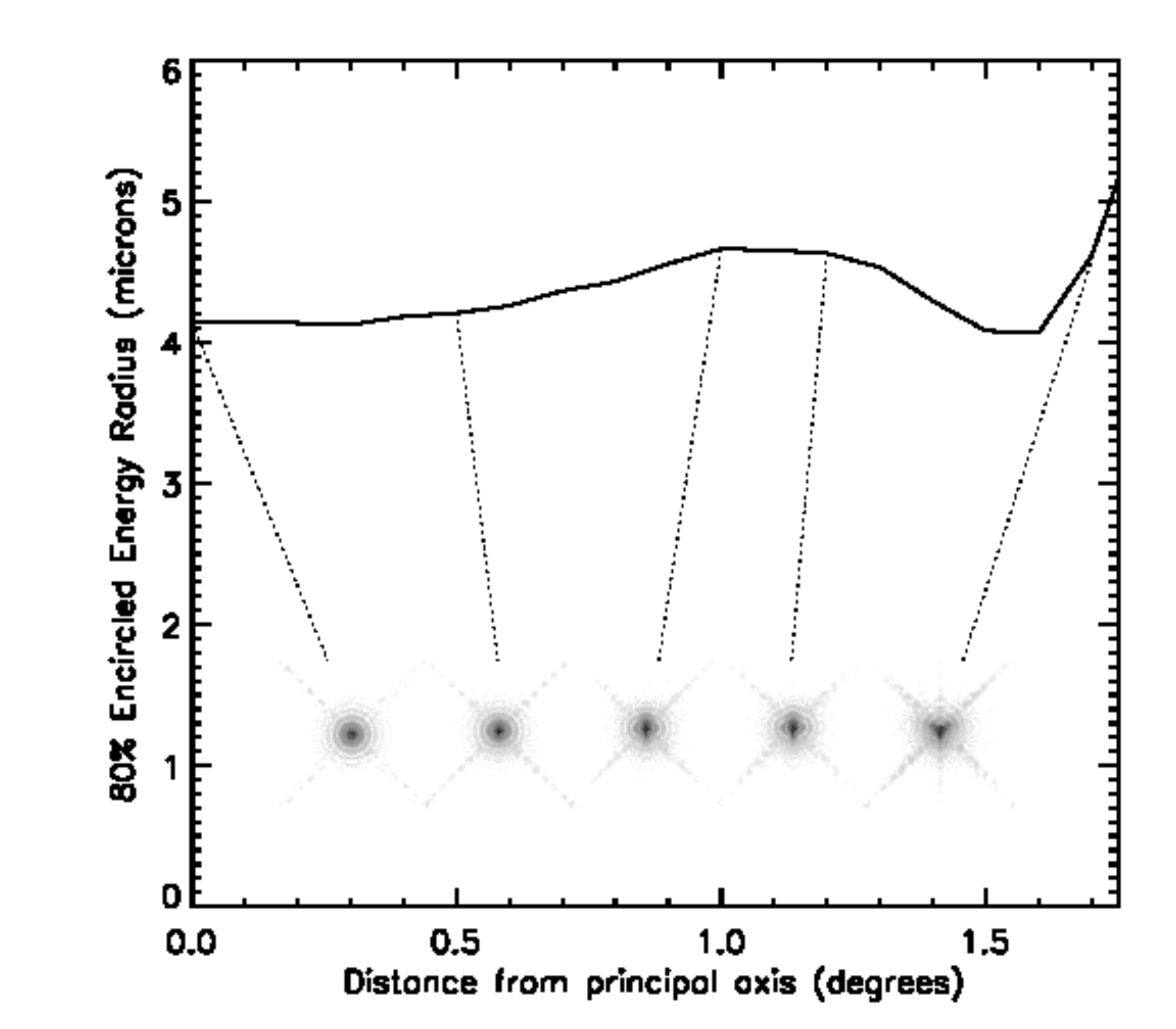}
\caption{Encircled energy of LSST PSF. Here we estimate the diffraction-limited encircled energy for $i$ filter.
The left panel shows the encircled energy as a function of radius for the four locations on the focal plane.
On the right panel the 80\% encircled energy radius as a function of the focal plane location (distance from
the principal axis) is plotted. The inset images display the snapshot of the PSFs at 0$\degr$ (center), 0.5$\degr$, 1$\degr$,
1.2$\degr$, and 1.75$\degr$ (edge).
Note the remarkably small variation of the PSF size across the entire focal plane. 
\label{fig_lsst_ee}}
\end{figure}

This optical design provides a wide field of view while a maintaining uniform image quality across the field (Seppala 2002). The remarkably small variation of the
PSF size over the field is illustrated in Figure~\ref{fig_lsst_ee}, where the diffraction-limited PSF images are generated with ZEMAX at the nominal
flat focal plane.
The peak-to-valley 80\% encircled energy radius is within $\sim7$\% of the
mean value from the center to $\sim1.4\degr$ for simulated $i$ filter point sources. The maximum deviation ($\sim17$\%) is seen 
at the edge of the field ($1.75\degr$).
The image brightness is stable across the field, providing a nearly uniform illumination within
a radius of 1.2$\degr$ and about 10\% decrease at the edge (Seppala 2002; 	LSST Science Collaborations 2010). 
Finally, the LSST field distortion is remarkably small: less than 0.1\% over the full field.

LSST will have continuous correction of its optics, using curvature wavefront sensing. Four special purpose rafts, mounted at the corners of the
science array, contain wavefront sensors and guide sensors (right panel of Figure \ref{fig_lsst_optics}).  Wavefront measurements are accomplished using curvature sensing,
in which the spatial intensity distribution of de-focused stars is measured at equal distances on either side of focus. Each curvature sensor is composed 
of two CCD detectors, with one positioned slightly above the focal plane, the other positioned slightly below the focal plane. 
The CCD technology for the curvature sensors is identical to that used for the science detectors in the focal plane, except that the curvature 
sensor detectors are half-size so they can be mounted as an in-out defocus pair.  Detailed analyses (Manuel et al. 2010) have verified that this configuration can reconstruct the wavefront within the $\lesssim0.2\mu$m accuracy.
These four corner rafts also hold two guide sensors each.  The guide sensors monitor the locations of bright stars at a frequency of $\sim10$ Hz to provide feedback for a loop 
that controls and maintains precision tracking of the telescope during an exposure. 

\begin{figure}
\includegraphics[width=8cm]{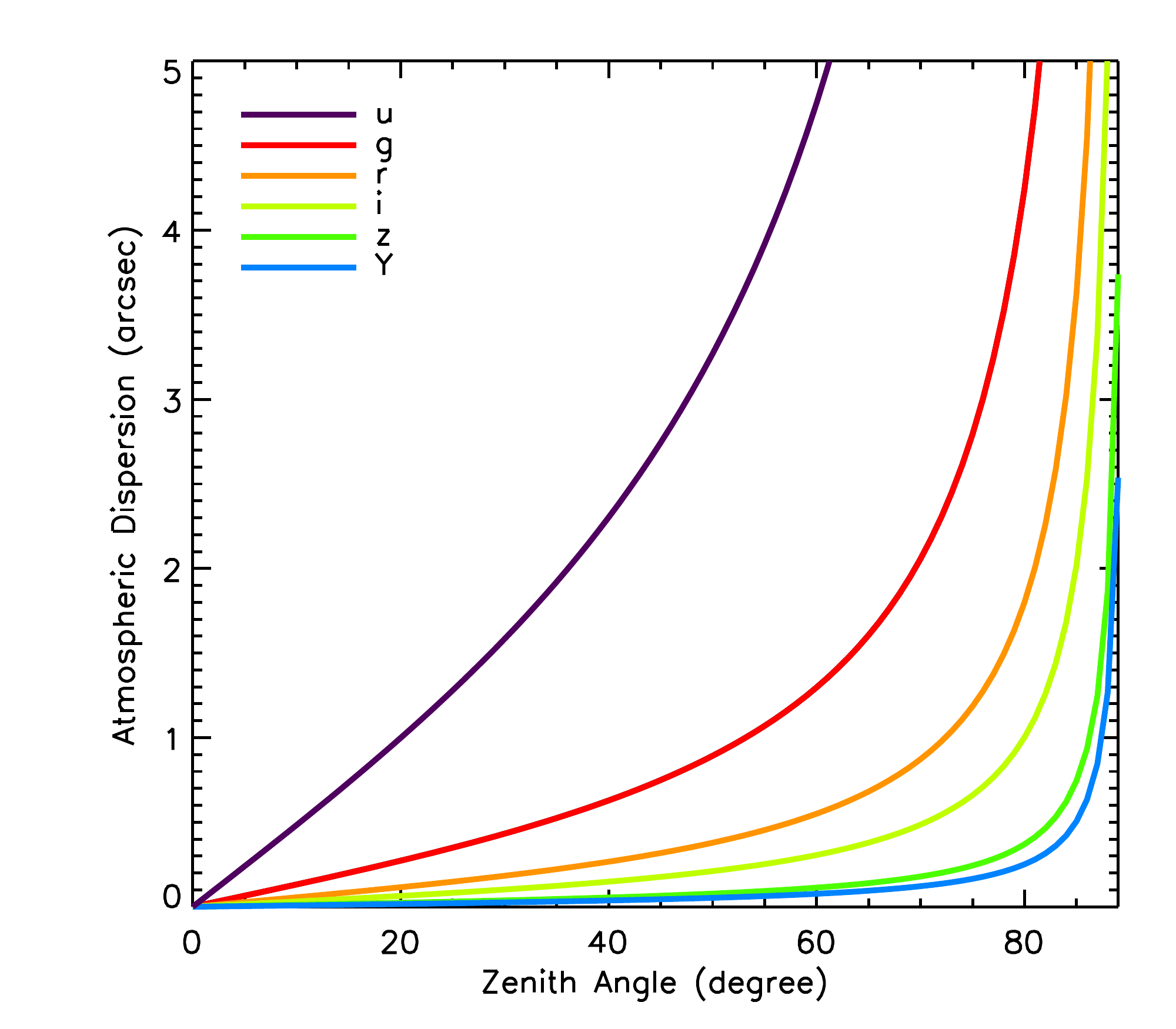}
\caption{Simulated Atmospheric dispersion at the proposed LSST site. We used the dispersion models summarized in Filippenko (1992) assuming
$f=8$ mm Hg (water vapor pressure), $T=5\degr$C (atmosphere temperature), and $P=520$ mm Hg (atmospheric pressure), which are
the approximate average conditions at Cerro Pach{\'o}n (Claver et al. 2004).
\label{fig_atmospheric_dispersion}}
\end{figure}

The fast focal ratio and the rapid pointing changes make any hardware-based atmospheric dispersion correction technically difficult.
Consequently, the effect of the atmospheric dispersion sets the maximum angle away from the zenith. We estimate the
atmospheric differential refraction for the six proposed filters ($u,g,r,i,z,$ and $Y$) using the models summarized in Filippenko (1982).
Figure~\ref{fig_atmospheric_dispersion} displays the results for the input parameters of $f=8$ mm Hg (water vapor pressure), $T=5\degr$C (atmosphere temperature), and
$P=520$ mm Hg (atmospheric pressure), which are the approximate average conditions at Cerro Pach\'{o}n (Claver et al. 2004).
Because the weak-lensing shape measurements will be carried out in $r$ and $i$ band images, it is critical for the $r$ and $i$ band imaging to provide the excellent seeing
while still covering the required sky area of 20,000 square degrees. 
From Cerro Pach\'{o}n, the required 20,000 square degrees of survey area are viewable above a zenith angle of 45 degrees.
At this 45$\degr$ limit, the corresponding dispersion in the $r$ and $i$ bands is less than $\sim0.3\arcsec$, which elongates the PSF in the zenith angle direction by $\sim9$\% and gives a coherent PSF ellipticity of $\lesssim0.04$ for $\sim0.7\arcsec$ seeing.
This amount of ellipticity by atmospheric dispersion is small and comparable to the values induced by the optical aberrations (after atmospheric turbulence). Therefore, we conclude
that the LSST design does not require atmospheric dispersion corrector (ADC).

\section{IMPLEMENTATION OF ATMOSPHERIC TURBULENCE}
 \label{section_atmosphere}
After the optics and camera are corrected by the wavefront sensing, the largest residuals are due to the atmosphere and the focal plane non-flatness.
The simulation of the atmospherically distorted wavefront is the first step for the generation of the realistic PSF of LSST.
A typical short exposure image of a star through atmospheric turbulence can be viewed as a superposition of a number of
speckles. The spatial extent of each speckle is roughly determined by the diffraction limit of the instrument whereas
the relative centroid shift between these speckles reflects the severity of the turbulence mainly depending
on the amplitude of the low frequencies. To mimic these effects numerically, we follow the well-known classical approach which 
regards the atmosphere as consisting of multiple moving layers each with 
independent phase screen and velocity.
Although a single phase screen has been shown to provide many characteristics of the real environments qualitatively,
the wind speed must be unrealistically high $\sim100~\mbox{km~s}^{-1}$ in order to correctly represent the time scale of the observed speckle
changes. 
In principle, the turbulence affects both the magnitude and phase of the wavefront. However, because
the phase distortion has a much larger effect for non-saturated scintillation (Fried 1994), in this paper we only consider the phase effect on the PSF behavior.

The power spectrum of the phase screen is usually described by the Kolmogorov/von K{\'a}rm{\'a}n model (Kolmogorov 1941; von K{\'a}rm{\'a}n 1930; Sasiela 1994):
\begin{equation}
\Psi (\nu)=0.023 r_0^{-5/3} \left ( \nu^2 + \frac{1}{L_0^2} \right )^{-11/6},
\end{equation}
\noindent
where $\nu$, $r_0$, and $L_0$ are the two-dimensional frequency, the Fried parameter, and the outer scale, respectively.
The outer scale $L_0$ is infinite for Kolmogorov screens. However, for a large
telescope this approximation becomes invalid, leading to an over-prediction of the PSF size (e.g., Tokovinin \& Travouillon 2006). 
We set
$L_0$ to 25 m, which is close to the average value at Cerro Pach{\'o}n determined by Ziad et al. (2000) from the measurement
of the angle-of-arrival fluctuations.

The realization of the Kolmogorov/von Karman phase screen in the spatial domain is obtained by taking the inverse Fast Fourier Transform (FFT) of
$\psi (\nu)$, whose modulus is given by $L\sqrt{\Psi(\nu)}$; $L$ refers to the physical length of the screen and
the phase of $\psi (\nu)$ is random.
For the two-dimensional array, where $i$ and $j$ are the indexes
in the spatial domain, and $k$ and $l$ are the indexes in the frequency domain, $\psi(i,j)$ (the phase value at $i$ and $j$) is given by (e.g., Carbillet \& Riccardi 2010)
\begin{eqnarray}\nonumber
\psi (i,j)&=& \sqrt{2} \sqrt{0.023} \left ( \frac{L}{r_0} \right )^{5/6} \times  \\
 &&   \mbox{FFT}_{R/I}^{-1} \left [ \left ( k^2+l^2+\left ( \frac{L}{L_0} \right )^2 \right )^{-11/12} 
e^{\phi (k,l) \bvec{i}} \right ].   \label{eqn_phase_screen}
\end{eqnarray}
\noindent
In equation~\ref{eqn_phase_screen}, $\mbox{FFT}_{R/I}^{-1}$ means that either real or imaginary part of the inverse FFT result is
used (thus the additional factor $\sqrt{2}$ is present). $\phi(k,l)$ is the uniformly distributed random number in the range $-\pi\leq\phi<\pi$.
Because we used FFT, one obvious drawback of the phase screen generated in this way is the lack of frequencies lower than the value
set by the physical dimension of the array (i.e., $1/L$). Lane et al. (1992) suggest a compensation of this artifact by
adding subharmonics to the result. However, this problem is not relevant in the current study because the simulation of LSST
requires a large ($\gtrsim$ 1 km) phase screen to account for the large field of view. 

\begin{figure}
\includegraphics[width=12cm]{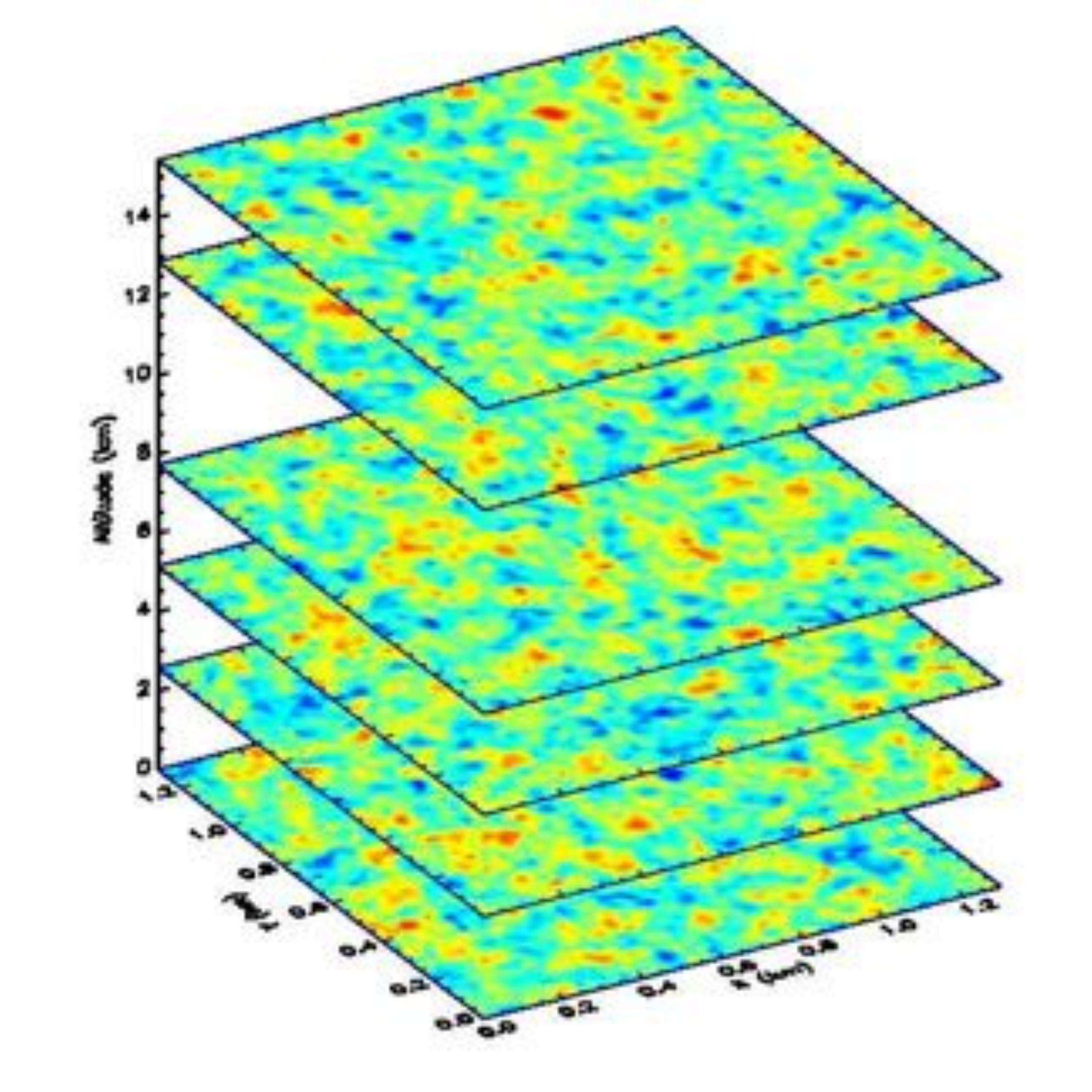}
\caption{Six layers of Kolmogorov/von Karman phase screens used for the atmospheric turbulence model.
\label{fig_phase_screen}}
\end{figure}

For the realization of the specific environment for LSST, we adopt the results of Ellerbroek (2002), 
who provides the parameters of the atmospheric turbulence profile based on the measurements at Cerro Pach\'{o}n, Chile, the
site of the current Gemini-South telescope and also the future LSST observatory. 
The atmospheric profile consists of six layers 
at altitudes of 0, 2.58, 5.16, 7.73, 12.89, and 15.46 km as shown in Figure~\ref{fig_phase_screen}  with 
relative weights of 0.652, 0.172, 0.055, 0.025, 0.074, and 0.022, respectively (Ellerbroek 2002).
The Fried parameter ($r_0$) is set to 0.16 m, which is reported to be an approximate median condition at the location by Ellerbroek (2002);
for a large telescope with a diameter $D>>r_0$ short-exposure seeing 
is inversely proportional to $r_0$.
Each layer is allowed to move independently at a constant velocity during the integration. The maximum wind velocity at the highest altitude 
is kept under $\sim20 ~\mbox{m}~\mbox{s}^{-1}$,
which determines the minimum time step 0.005 s (corresponding to $\sim0.5$ pixel shift) to satisfy Nyquist sampling rate.
We choose the dimension of our resulting phase screen array to be sufficiently large ($6\times8192\times8192$ or $6\times 1.3 \mbox{km}\times 1.3 \mbox{km}$) in order to keep the moving phase screens covering the telescope field of view during the 15 s integration time.

\begin{figure}
\includegraphics[width=18cm]{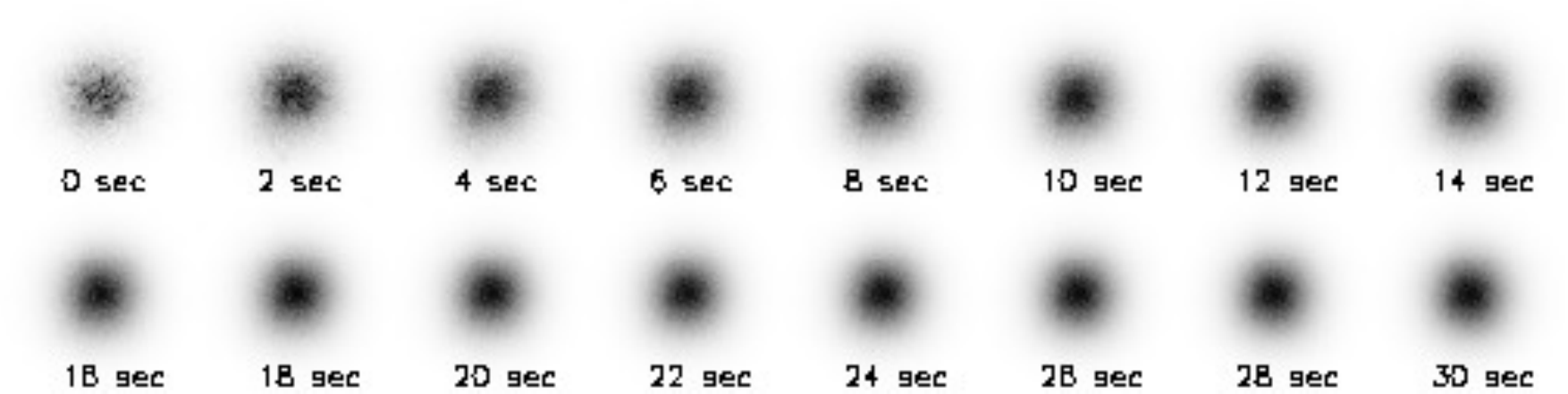}
\caption{Example of time evolution of simulated LSST PSF due to atmospheric turbulence at the zenith.
The displayed PSFs are for the monochromatic beam at $\lambda=6000$\AA.
We do not include the optical aberration or the charge diffusion by CCDs in this simulation. 
As the exposure time increases, more speckles are stacked together, which makes the resulting PSF rounder and the surface brightness irregularities
decrease. The FWHM at $t\sim15$s is $\sim0.5\arcsec$.
\label{fig_psf_time_evolution}}
\end{figure}

The weighted average $\psi(x,y)$ of the six phase screens at a given moment can be converted to a snapshot of the telescope PSF 
in the absence of the optical aberration via the following equations:
\begin{equation}
p(x,y)=A(x,y) e^{ i 2\pi \psi (x,y) /\lambda } \label{eqn_p}
\end{equation}

\begin{equation}
PSF=| FFT(p) |^2, \label{eqn_psf}
\end{equation}
\noindent
In equation~\ref{eqn_p}, $A(x,y)$ is the telescope pupil function (a mask showing the obscuration)
and the phase difference $\psi(x,y)$ has the dimension of length.
The pixel scale of the PSF array given by equation~\ref{eqn_psf} is simply $F \lambda/2$, where $F$ is the f-ratio.
Figure~\ref{fig_psf_time_evolution} displays the examples of the LSST PSFs generated in this way at different integration times. The PSF at $t=0$ shows 
the typical instantaneous speckle. In this figure we do not include either the optical aberration or the charge diffusion by CCDs, the effects of which
we however later add to generate the simulated LSST images.
As the exposure time increases, more speckles are stacked together, which makes the resulting PSF rounder and the irregular features present in the
individual speckles more smeared (de Vries et al. 2007).

A quantitative study on the impact of the atmospheric turbulence on the ellipticity and its spatial correlation is needed to support the validity of
our simulation hereafter. As is discussed in \textsection\ref{section_focal_plane}, we model LSST PSF variation CCD-by-CCD with polynomials.
If the anisotropic power from the atmosphere within 15 sec exposure turns out to be too strong and changes on a very small scale, any conventional interpolation
scheme with the finite number of stars will over-smooth the inherent PSF variation.

We define the ellipticity of PSF as $(a-b)/(a+b)$, where
$a$ and $b$ are the semi-major and -minor axes of the PSFs, respectively. 
In general, the PSF isophotes change with radius. Thus, the measurement somewhat depends on the weighting scheme.
We measure the ellipticity of the PSF using the following quadrupole moments:
\begin{equation}
Q_{ij} = \frac{ \int d^2 \theta W(\symvec{\theta}) I(\symvec{\theta}) (\theta_i - \bar{\theta_i})(\theta_j - \bar{\theta_j}) }
           {\int d^2 \theta W(\symvec{\theta}) I(\symvec{\theta}) }, \label{eqn_quadrupole}
\end{equation}
\noindent
where $I(\symvec{\theta})$ is the pixel intensity at $\symvec{\theta}$, $\bar{\theta}_{i(j)}$ is
the center of the star, and $W(\symvec{\theta})$ is the optimal weight function required to suppress the noise in the outskirts.
For the diffraction-limited PSF of LSST (Figure~\ref{fig_ellipticity_focal_plane_no_atmos}), we choose a Gaussian with a FWHM of  $\sim$0.3\arcsec for $W(\symvec{\theta})$ whereas for the turbulence-limited PSF the FWHM of the Gaussian weight function used is $\sim0.7\arcsec$.
The quadrupole moments are converted to the ellipticity $\symvec{\epsilon}$ via:
\begin{eqnarray}\nonumber
\symvec{\epsilon} &=& (\epsilon_+,\epsilon_{\times}) \\
\nonumber
& = & \left ( \frac{Q_{11}-Q_{22}} {Q_{11}+Q_{22}+2(Q_{11} Q_{22}-Q^2_{12})^{1/2}} , \right. \\
&& \left. \frac {Q_{12}} {Q_{11}+Q_{22}+2(Q_{11} Q_{22}-Q^2_{12})^{1/2}} \right ) \label{eqn_delta}.
\end{eqnarray}
\noindent
The magnitude of the stick is proportional to $|\symvec{\epsilon}|=\sqrt{\epsilon_+^2 + \epsilon_{\times}^2}$, and the orientation
angle is given by $0.5\tan^{-1}( \epsilon_{\times}/ \epsilon_{+})$.

\begin{figure}
\includegraphics[width=9cm]{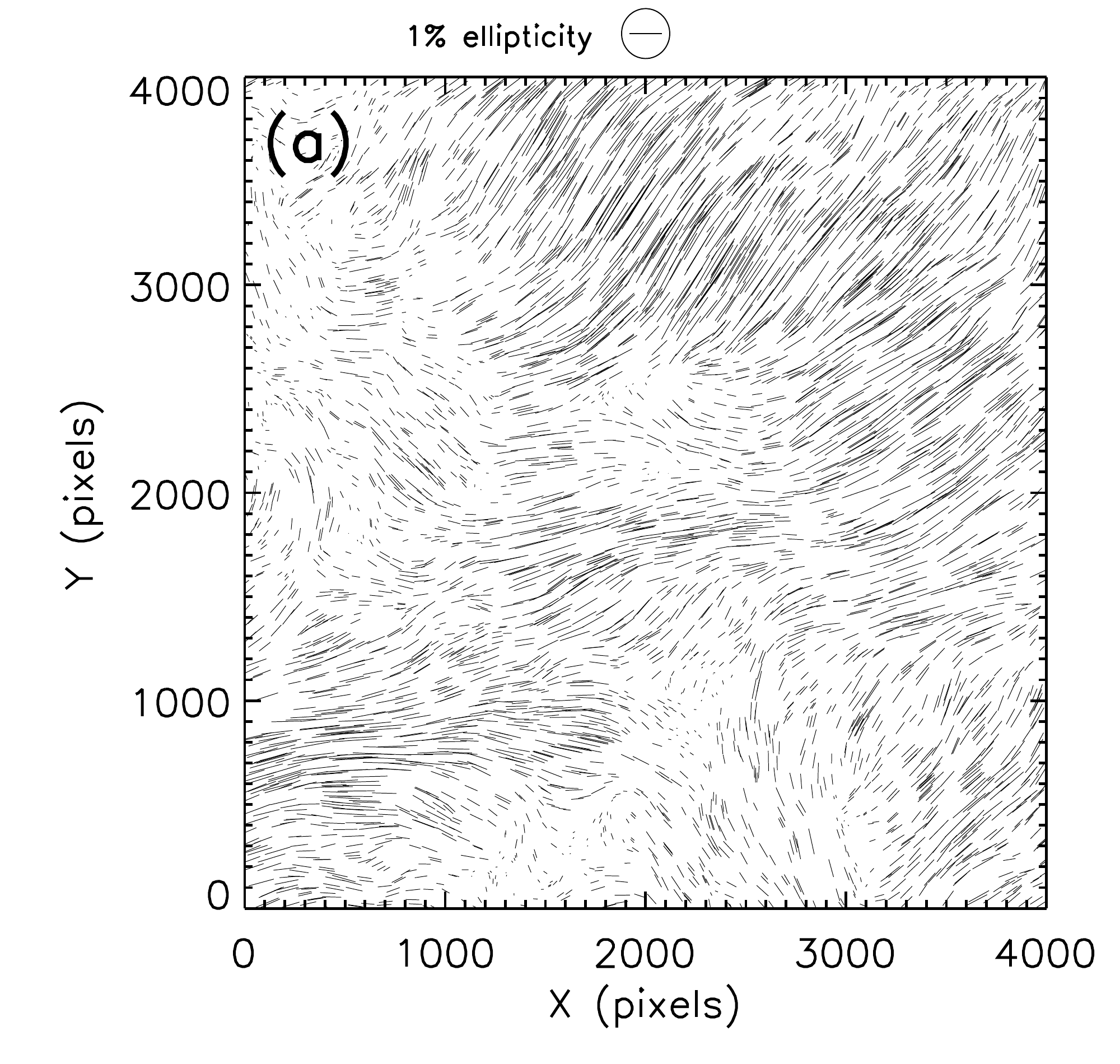}
\includegraphics[width=9cm]{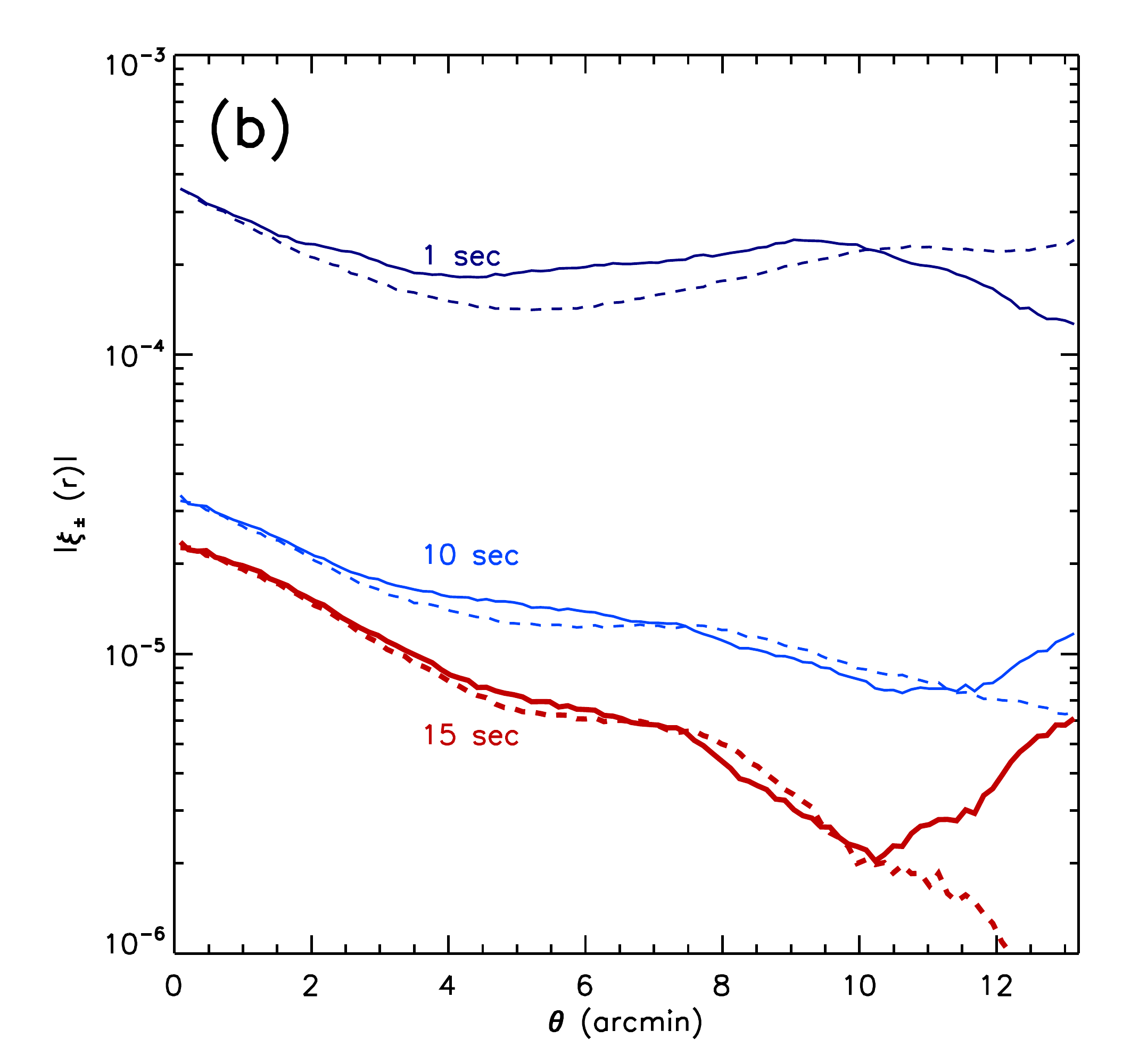}
\caption{Impacts of turbulence on PSF. (a)  PSF ellipticity (spatial) variation for the 15 s exposure within a 4k$\times$4k CCD of LSST due to atmosphere.  Because no optical aberration is introduced yet, this plot shows purely atmospheric contribution. A clear spatial correlation is visible on a scale of $\sim$1$\arcmin$ ($\sim$300 pixels). However, the average magnitude is less than 1\%, and thus the resulting anisotropy does not contribute significantly to the final PSF. 
(b) Ellipticity correlation for different exposure times. The amplitude of the correlation function decreases with exposure time because the telescope aperture sees higher number of (effectively) uncorrelated phase screens.  The solid and dashed lines represent $\xi_{+}$ and $\xi_{-}$, respectively.
\label{fig_at_psf}}
\end{figure}

Figure~\ref{fig_at_psf}a displays the PSF ellipticity variation for the 15 s exposure within a 4k$\times$4k CCD of LSST by atmosphere.  Because no optical aberration is introduced yet, this 4000 ``whiskers'' show the purely atmospheric contribution. 
The size of the sticks represents the magnitude of the ellipticity whereas the
the orientation of the stick is aligned with the position angle of the elongation. A clear spatial correlation is visible on a scale of $\sim$1$\arcmin$ ($\sim$300 pixels). However, the average magnitude is less than 1\%, and thus the resulting anisotropic power is not strong. Figure~\ref{fig_at_psf}b shows the
ellipticity correlation based on the following equation:
\begin{equation}
\xi_{\pm}(r)=\langle e_{t} (r^{\prime} )~ e_{t} (r^{\prime} + r) \pm  e_{\times} (r^{\prime} )~ e_{\times} (r^{\prime} + r) \rangle,
\end{equation}
\noindent 
where the product $e (r^{\prime} )~ e (r^{\prime} + r)$ is performed for each pair separated by $r$, and 
the $e_{t}$ and $e_{\times}$ are  the tangential and the 45$\degr$ components,respectively, with respect to the line
connecting the pair. 
The amplitude of the correlation function decreases with exposure time.  From the comparison with the contribution due to the telescope and camera aberration 
(see \textsection\ref{section_focal_plane}), we conclude that the small scale anisotropic power for the delivered PSF is not dominated by the atmospheric
turbulence. These results have been validated with a series of 15 sec exposures of the globular cluster NGC2419 on the Subaru telescope.

\section{LSST FOCAL PLANE AND IMPACTS ON PSF}
\label{section_focal_plane}
\begin{figure}
\includegraphics[width=8cm]{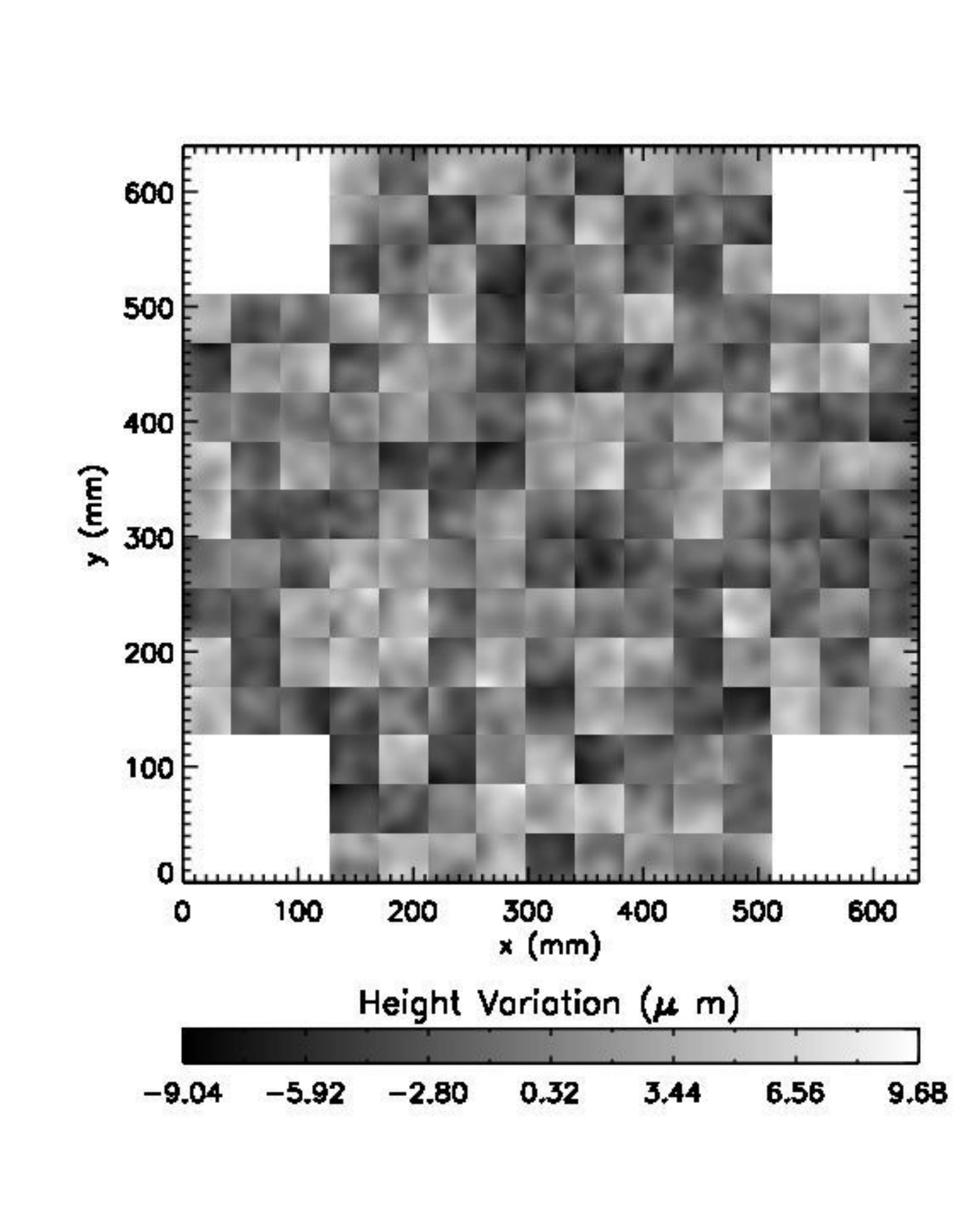}
\caption{Simulated focal plane of LSST. We used the CCD assembly/fabrication
specification in Table 1 to generate the LSST focal plane tiled by 189 CCDs.
\label{fig_focal_plane}}
\end{figure}

\begin{figure}
\includegraphics[width=18cm]{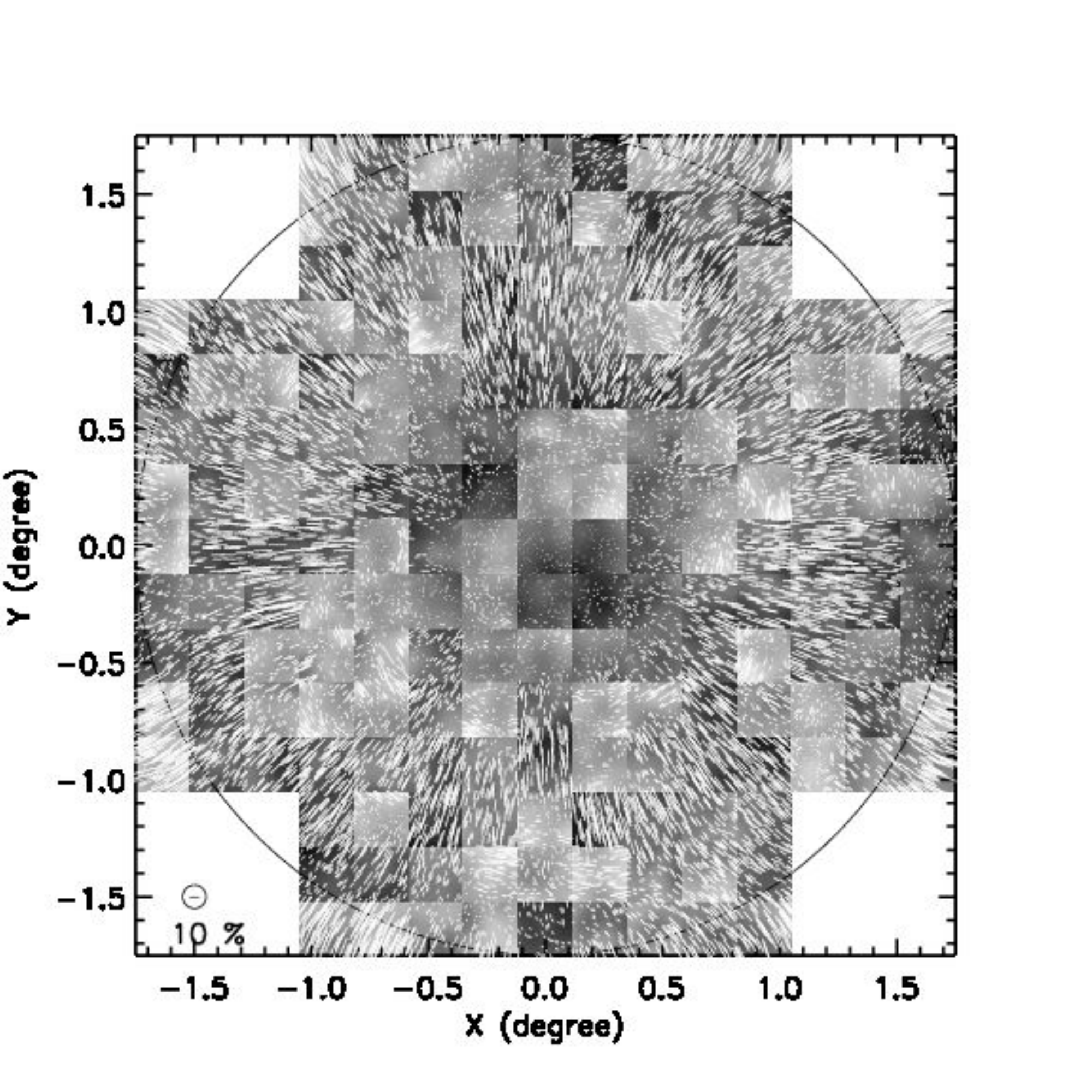}
\caption{Impact of focal plane CCD height variation on PSF. We use the ZEMAX software to obtain diffraction-limited PSFs in the absence of atmospheric turbulence.
The encircled stick in the lower-left corner represents 10\% ellipticity as illustrated. Note both the magnitude of ellipticity induced by
the focal plane error (and aberration) and the abrupt changes across the CCD borders. The large circle shows the 3.5 m diameter field of view.
The sticks outside this circle are for illustrative purpose only to demonstrate that the aberration degrades severely beyond the 1.75 $\degr$ boundary.
\label{fig_ellipticity_focal_plane_no_atmos}}
\end{figure}

The 64 cm diameter focal plane of LSST will be tiled with 189 4k$\times$4k CCDs. As-built heights might vary up to $\sim10$ microns (peak-to-peak) in a complicated way
relative to the nominal flat surface. Although this flatness deviation is small compared with other cameras in existing facilities, the small f-ratio
of the instrument makes this focal plane flatness variation play a critical role in aberration-induced PSF behavior (depth of focus $\propto$ f-ratio). 
In Table 1 we summarize the current specification of the focal plane error budget originally issued to vendors, as well as the values used for the current simulation. The CCD flatness technology and testing is reviewed in Takacs et al. (2006) and Radeka et al. (2009).  The focal plane errors that we assume for the simulations in this paper are larger than the current expectation so that the results presented here are conservative. Several vendors are now exceeding these flatness specifications.
A realization of the LSST focal plane based on these error distributions
is displayed in Figure~\ref{fig_focal_plane}.
The resulting behavior of the PSF ellipticity is illustrated in Figure~\ref{fig_ellipticity_focal_plane_no_atmos}. These PSFs are obtained by evaluating optical path difference
functions across the focal plane in the absence of atmospheric turbulence using the ZEMAX software; see also Jarvis, Schechter, \& Jain (2008) for the discussion on the impact of the telescope focus on the PSF ellipticity.

Several features are noteworthy in Figure~\ref{fig_ellipticity_focal_plane_no_atmos}. First, the aberration-induced ellipticity is large. Most of the large sticks
in the plot exceed $\sim10$\% ellipticity, approaching nearly $30$\% at the field edges. 
Because the maximum deviation in this realization
is small ($\sim10$ $\mu$m), these large values of ellipticity remind us that the focal error tolerance of LSST is indeed narrow;
however, we note that in the central region ($\lesssim 1$\degr) the sensitivity of PSF elongation to height error is somewhat mitigated.
Second, sharp discontinuities of PSF ellipticity are present where the CCD heights also vary abruptly.
Third, although it may be a bit difficult to recognize this feature in Figure~\ref{fig_ellipticity_focal_plane_no_atmos}, a small-scale variation is 
observed even within a CCD mainly due to the
tilt and potato chip effects.

The presence of these smooth, small-scale variation and discontinuous changes across CCD gaps are the most challenging aspects of the PSF description and modeling
for LSST. Any attempt to use a single set of polynomials to characterize the PSF behavior across the entire focal plane fails because the small-scale variation
requires impractical, high-order terms in the polynomials, and in addition no interpolation scheme can satisfactorily 
reproduce the sharp discontinuities across the CCD borders. This is the reason that in the current paper we perform interpolation CCD-by-CCD
to model LSST PSFs. Considering the 1-2 cycles of the potato chip effect within a CCD, we estimate that 3-4 order polynomials suffice. 
The remaining
question is what features of PSF are interpolated. A simple description of the two components of the PSF ellipticity has been used in the early weak-lensing
analysis (e.g., Valdes, Jarvis, \& Tyson 1983; Kaiser, Squires, \& Broadhurst 1995). However, certainly ellipticity alone does not fully characterize a PSF. One obvious missing piece of information is the size of the PSF.
One can imagine that a 10\% ellipticity PSF with FWHM=$1\arcsec$  more strongly affects the shape of galaxies than a PSF with the same
ellipticity but with FWHM=0.5$\arcsec$. Many authors (e.g., Hoekstra et al 1998, Kaiser 2000; Rhodes, Refreiger, \& Groth 2001)
suggested modifications to this earlier method (Kaiser, Squires, \& Broadhurst 1995) for the treatment of the PSF size.
Nevertheless, this increased sophistication (i.e., ellipticity with size) is not still sufficient to obtain the accuracy that the LSST weak-lensing science requires.

On the other extreme, one might consider interpolating pixel values of observed PSFs. The result of this type of PSF interpolation is
the actual image of the PSF, which in principle contains the most complete information at any point in the focal plane. 
However, this scheme faces severe practical challenges in the implementation because
1) except for the pixels near the core, individual PSF images are noisy, and 2) the number of needed polynomials become prohibitively large as one enlarges
the dimension of the PSF. Therefore, one might imagine that a new PSF description scheme should provide sufficient details with both an efficient
noise filtering method and a compact set of polynomials. A natural choice is the description of the PSFs using some basis functions, preferably
an orthogonal set to avoid degeneracy and thus improve compactness. Bernstein \& Jarvis (2002) and Refregier  (2003) suggest the use of eigenfunctions of 
two-dimensional quantum harmonic oscillator (also known as shapelets) for this purpose. Although the shapelet approach is shown to work 
in many applications, Jee et al. (2007) demonstrate that the basis functions are still sub-optimal, truncating the PSF profile much earlier than desired.
Consequently, Jee et al. (2007) proposed to derive the basis functions from the data themselves via principal component analysis (PCA).

Lupton et al. (2001) first mentioned the use of PCA for the modeling of the Sloan Digital Sky Survey (SDSS).
Although the details of the implementation were not given, the method is similar to that of Jee et al. (2007) in that
PCA is employed to derive the optimal basis functions from the data themselves. On the other hand,
the use of PCA by Jarvis \& Jain (2004) and Schrabback et al. (2010) for the description of the PSF variation is somewhat different 
from the method of Jee et al. (2007) because they use PCA to characterize the pattern of the variation (e.g., coefficients or ellipticity components),
not to capture the pixelated PSF features.

Our PCA approach has been so far applied to weak-lensing studies only with Hubble Space Telescope images (e.g., Jee et al. 2009). 
In the current investigation, we
adopt this PCA scheme to characterize/model the LSST PSFs. As our PCA algorithm is independent of the specification of the instrument, 
the implementation in Jee et al. (2007) can be used with little modifications. 
The technical details of the PCA method will be reviewed in the following section.

\section{PRINCIPAL COMPONENT ANALYSIS }
\label{section_pca}

PCA is the mathematical procedure for analyzing multivariate data by finding and utilizing a new orthogonal coordinate system, which
substantially reduces the number of axes relative to the original number of variables for the characterization of the data.
Inevitably, the procedure involves information loss. However, in the typical data set where the noise is significant, this
reduction of the information in fact works as filtering, enabling a compact description of the data without significantly harming its integrity.
In the current paper, we explain the procedure
using our specific data set, i.e., series of noisy two-dimensional PSF images. A similar account can be also found in Jee et al. (2007).

Let us assume that we have $N$ observed stars in a 4k$\times$4k CCD image with random positions and brightness. 
The two-dimensional
array of each PSF image can be rearranged to form a one-dimensional vector. How we exactly scramble the pixels of the PSF image
is not important in the subsequent analysis as long as we have the ability to re-order the pixels to recover the original two-dimensional image.
However, it is critical that each PSF image is carefully background-subtracted and normalized prior to the rearrangement.
Then, the $N$ one-dimensional vectors are stacked to comprise a matrix. If each PSF image contains $M$ pixels, the result is
a $N\times M$ matrix. The next step is to evaluate mean values along the columns, and subtract this $1\times M$ mean vector
from each row of the matrix. In practice, each row of the $N\times M$ matrix contains star images contaminated by neighboring
objects, cosmic rays, bad pixels, etc. Therefore, the mean values are preferably evaluated iteratively with some outlier rejection.
Also, the rejected pixel values in the $N\times M$ matrix should be replaced with the mean values.
Because we need this mean vector to reconstruct the PSF later, this mean vector (or mean PSF rearranged
in one-dimension) must be saved. 

Now, we are ready to find a set of $P$ ($\ll MIN\{N,M\}$) orthogonal vectors that compactly describe the $N\times M$ matrix. One way to
derive such orthogonal vectors is Singular Value Decomposition (SVD; Press et al. 1992). If we call the $N\times M$ matrix \bvec{S},
the matrix \bvec{S} can be decomposed in the following way:
\begin{equation}
\bvec{S}=\bvec{U}\bvec{W}\bvec{V}^T \label{eqn_SVD},
\end{equation}
\noindent
where $\bvec{U}$ is an $N\times M$ column-orthogonal matrix, $\bvec{W}$ is an $M\times M$ diagonal matrix, and
$\bvec{V}$ is an $M\times M$ orthogonal matrix. The diagonal elements of $\bvec{W}$ are called singular values.

\begin{figure}
\includegraphics[width=18cm]{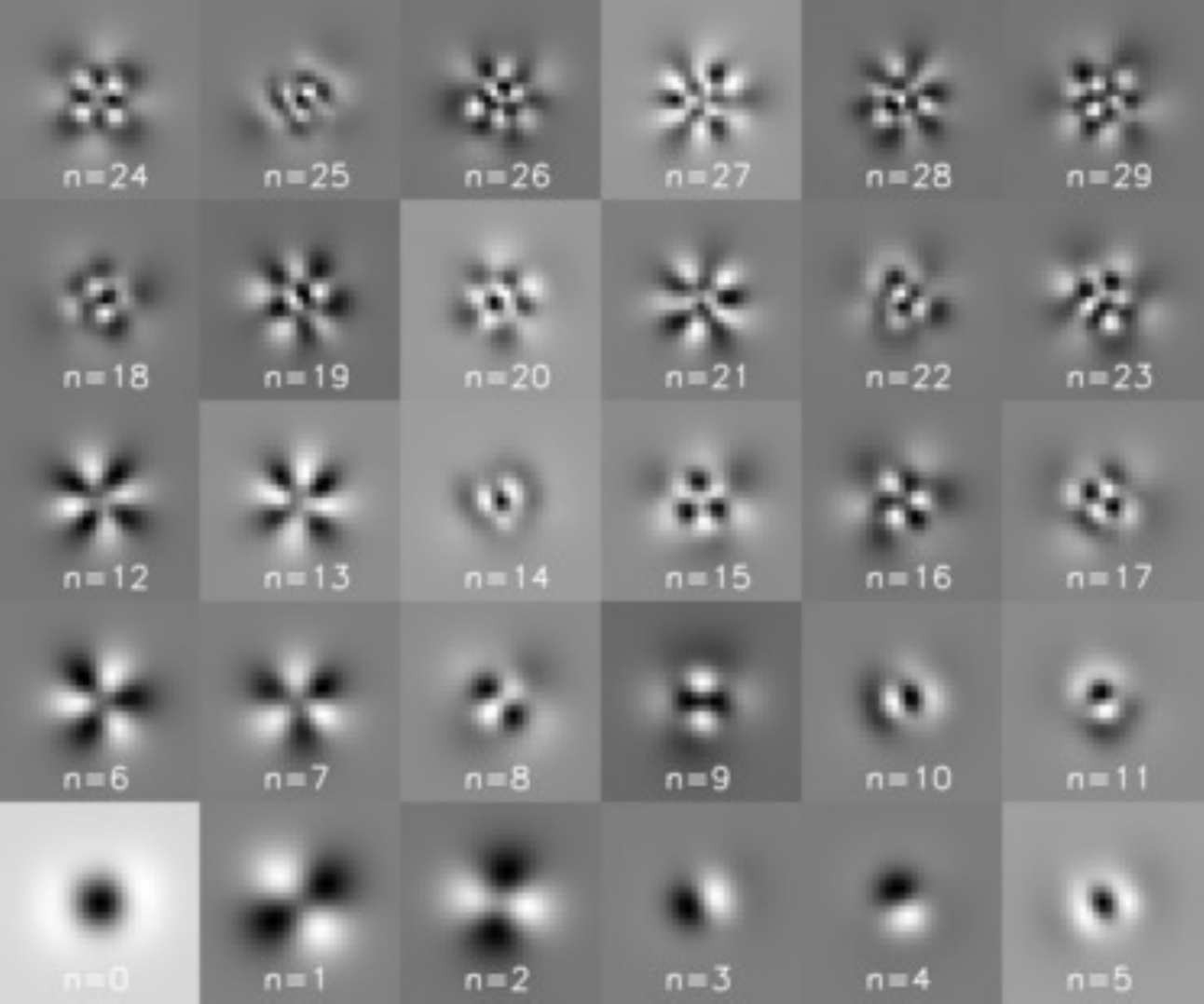}
\caption{PCA eigenPSFs. The eigenPSFs were ranked by their eigenvalues with $n=0$ representing the largest.
The example shown here is derived by performing PCA on 4000 simulated PSFs. As $n$ increases, the eigenPSF
tends to possess higher-frequency features. Interestingly, the first several eigenPSFs resemble the basis functions
obtained by the shapelet formalism remarkably. However, asymmetry is also evident, which reveals the characteristic pattern of the data.
\label{fig_eigenPSF}}
\end{figure}

Therefore, the matrix component $\bvec{S}_{ij}$ can be expressed as
\begin{equation}
S_{ij} = \sum_{k=1}^{M} w_k U_{ik} V_{jk}. \label{eqn_svd_component},
\end{equation}
\noindent
which implies that, if some singular values $w_k$ are relatively small, we can substitute zeros for those $w_k$'s without greatly
compromising the numerical accuracy of the matrix evaluation. In other words, the reconstruction of $\bvec{S}$ is
possible with fewer columns in both $\bvec{U}$ and $\bvec{V}$, and thus the procedure provides a substantial compression
of the information in the original matrix $\bvec{S}$, where photon and detector noise is a significant part.
The remaining columns (principal components) of $\bvec{V}$ form an orthogonal basis, and in this analysis we refer to these eigenfunctions
as eigenPSFs. We display examples of these eigenPSFs in Figure~\ref{fig_eigenPSF}. These eigenPSFs in this figure are
obtained from the PCA of 4000 simulated stars on the entire focal plane in the absence of telescope aberration and CCD height fluctuation, but
in the presence of the atmospheric turbulence. The eigenPSFs are ranked according to the size of their eigenvalues with $n=0$ representing the
largest. Note the remarkable resemblance of the first several eigenPSFs to the basis functions in the shapelet approach. However, also
note the asymmetry of the eigenPSFs due to the characteristic pattern of the specific data set. For increasing $n$, more high-frequency
features are evident. Above a certain limit $n=n_c$, the resulting eigenPSFs start to reflect the noise features without contributing 
to the PSF reconstruction. Consequently, we need only a truncated set of eigenPSFs to describe the behavior of the PSF.

This minimal number of eigenPSFs can be determined by examining the eigenvalues of each eigenPSF. In general, there
is a conspicuous ``kink'' at a certain $n=n_c$, beyond which the decrease of the eigenvalue suddenly slows down
(e.g., Figure 5 of Jee et al. 2007). We find that $\sim20$ eigenPSFs are sufficient to describe the details for both
space- and ground-based telescope PSFs.

After the eigenPSFs are obtained, the $k^{th}$ star used for the derivation is decomposed as
\begin{equation}
C_k (i,j)= \sum_{n=0}^{n_{max}} a_{kn} P_n (i,j) + T(i,j),
\end{equation}
\noindent
where $C_k (i,j)$ is the normalized pixel value of the $k^{th}$ star at the
pixel coordinate $(i,j)$,
$P_n$ is the $n^{th}$ eigenPSF, $a_{kn}$ is
the projection of the $k^{th}$ star in $P_n$, and $T$ is the mean PSF. Because $P_n$'s are orthogonal to
one another, one does not need to determine $a_{kn}$ through some $\chi^2$ minimization here.

Now the PSF at any arbitrary location can be evaluated through polynomial interpolation of $a_{kn}$'s:
\begin{equation}
a_n = \sum_{l,m=0}^{l+m\leq N} d_{nlm} x^l y^m
\end{equation}
\noindent
where $x$ and $y$ are the coordinate of the position, at which one wants to know the PSF, $d_{nlm}$
is the coefficient of the term $x^l y^m$ for the $n^{th}$ eigenPSF, and $N$ is the maximum
order of the polynomial. It is important to determine $d_{nlm}$ without the application
of the S/N weight of the star because we are interested in the spatial variation
of the PSF.

\section{SIMULATION OF LSST IMAGES AND PSF RECONSTRUCTION}
\label{section_simulation}
For the robust simulation of the weak-lensing shear extraction using LSST, 
it is crucial that the simulated image contains astronomical objects with
realistic photometric, morphological, and statistical properties, as well as the instrumental artifacts resulting
from the known characteristics of the telescope, camera, and the environment.
Specifically, the requirements include our ability to correctly represent:
\begin{enumerate}
\item the number density and luminosity function of unsaturated high S/N Galactic stars,
\item the number density, luminosity function, size, and morphology of galaxies,
\item position-dependent PSF variations,
\item geometric distortion, and
\item atmospheric turbulence and dispersion.
\end{enumerate}
The first requirement is needed because the number of available high S/N stars
determines how well we can measure the behavior of the PSFs. In particular, we need
to investigate if the nominal 15 s exposure image provides sufficient number of usable stars
within a CCD so as to correctly sample the position-dependent PSF variation.

The next point relates to the issue of both statistical and systematic errors
of the shear measurement from galaxy ellipticity. The number of usable galaxies
for testing lensing analysis affects the statistical accuracy of the estimate of shear and our determination of mass.
In addition, one important source of bias in many ellipticity measurement techniques is
the under-fitting of galaxy shapes (i.e., the model galaxy profile assumed is different
from the real profile). Since many $z\gtrsim1$ galaxies are dominated by
so-called faint blue galaxy (FBG) population incapable of being described by
analytic functions, it is important that the image contain a realistic fraction of
these irregular galaxies.	
The extremely high sensitivity of the LSST PSF to the focal plane errors makes the position-dependent
PSF one of the most critical features to be simulated. 
We need to	generate the realistic PSF and convolve the galaxy image with this position-dependent
PSF in a time-efficient way.

Although expected to be very small, the geometric distortion by the telescope and atmosphere 
can correlate ellipticity of galaxies and mimic gravitational lensing signal similar to the way that
uncorrected PSF can cause systematic alignment of galaxy ellipticity. It is essential to
estimate the level of accuracy in rectifying the raw images and the effects on shear measurement. 						

The atmospheric turbulence and dispersion are dominant sources of time-dependent PSF variation.
Unlike the telescope and camera optical aberration, the atmospheric turbulence is believed to circularize
the PSF rather than to introduce significant anisotropy whereas the atmospheric dispersion obviously
elongates the PSF in the direction of the zenith angle. It goes without saying that these atmospheric
effects should be properly included in the simulation.

Our simulation of the LSST shear test is an on-going project, and the final goal is to carry out an
end-to-end simulation from the realization of the LSST image from an input cosmology to the
full recovery of the galaxy shear signal and the cosmological parameters. Hence, it is our ultimate aim to rely on the
final simulator not only to implement the above requirements but also to address other non-shear measurement 
issues such as diffuse components, cosmic rays, hot/warm/bad pixels, etc.
This end-to-end simulation is being done by The LSST Simulations Team (Connolly et al. 2010).

Here, as a pilot study, we focus on generating realistic LSST PSFs and recovering these input
PSFs from sparse distribution of the stars in the presence of background galaxies. Therefore, some simplifications
are made where they do not significantly compromise the integrity of our simulation relevant to the focal plane
flatness issue.

\subsection{Generation of Input PSF Model }
 \label{section_inputpsf}
First, we generate the coordinates of the model stars that are densely distributed across the focal plane.
Because we will evaluate the fiducial PSFs at these locations and interpolate the results to describe the PSF
on every pixel, the density of these stars should be high enough to accommodate the expected small scale
variation. We assigned about 800 stars to each CCD and verify that these 800 stars sample with ample resolution the
spatial variation, which is dominantly modulated by potato-chip effects (1-2 cycles per CCD).
Then, we evaluate the diffraction-limited PSF at the location of those stars using ZEMAX. Strictly speaking,
in this step we save phase data rather than PSFs because the optical phase data should be later combined with
the atmospheric phase screen to generate the final PSF. We use the simulated focal plane height fluctuation shown in Figure~\ref{fig_focal_plane}
in order to provide the focus value for each star.  In the current investigation, our main
interest is the ability to model/recover the PSF in the $r$ band. Thus, we simulate the broadband effect by obtaining the data at five uniformly spaced wavelengths
between 5450-7030 \AA.
Next, the atmospheric phase screen is evaluated at the location of each star. This atmospheric phase information is
merged with the optical data to construct the next stage PSF. Finally, the result is further convolved twice 
to mimic the effects of the charge diffusion ($\sim0.4$ pixel rms) within a CCD and the atmospheric dispersion (close
to zenith).

We derive sets of eigenPSFs separately for each CCD. Alternatively, one can consider finding one set of eigenPSFs utilizing all PSFs on the focal plane. 
Although this latter scheme may reduce statistical errors, it appears that the resulting eigenPSFs are suboptimal, somewhat under-predicting
the systematic pattern unique in each CCD. We choose to keep the most signifincant 20 eigenPSFs, which accounts for
more than 99\% of the total variance. Now each PSF is described by 20 coefficients each representing the amplitude in each eigenPSF direction.
We fit 4th order polynomials to the coefficients of all stars to describe the PSF at any arbitrary location.

\subsection{LSST Galaxy Image Generation}
 \label{section_lsst_image}
One ideal way to generate galaxy images for a given cosmology is to make a catalog using large numerical simulation results (e.g., Millennium Simulation, Springel
et al. 2005) with
ray-tracing technique, to assign proper spectral energy distribution (SED) and morphology to individual galaxies, and to
convert the catalog to astronomical images. However, this full scale end-to-end simulation (Connolly et al. 2010) is beyond the scope
of the current project, and here we instead utilize Hubble Space Telescope (HST) Ultra Deep Field (UDF) images to sample
galaxies. The HST UDF (Beckwith et al. 2003) is a
412-orbit HST Cycle 12 program to image a single field with the Wide
Field Camera (WFC) of the Advanced Camera for Surveys (ACS) in
four filters: F435W, F606W, F775W, and F850LP. The total exposure time of
the F775W image is $\sim350$ ks, and the 10 $\sigma$ limiting magnitude is 29 ABmag. The depth of the UDF image
is sufficient for the generation of the galaxies that will be imaged by LSST, which reaches $r\sim 27.5$ mag ($\sim5\sigma$) in the final depth.
Hence, virtually, the UDF data provide noiseless, seeing-free galaxies (the size of the ACS PSF is less than that of one LSST pixel) for
a 15 s exposure.
As the field of view of the ACS is $\sim3\arcmin\times3\arcmin$, even smaller than the area covered by one LSST $\sim14\arcmin\times14\arcmin$ CCD,
it is inevitable that the same UDF galaxies must appear multiple times across the focal plane. However, this is not likely to compromise the
accuracy of the current weak-lensing simulation, where we do not attempt to recover any input cosmology. 

The high-level science product of the UDF is publicly available\footnote{http://archive.stsci.edu/prepds/udf/}, and we use the F775W image
to sample galaxies. Objects were detected via SExtractor (Bertin \& Arnouts 1996) by looking for 8 continuous pixels above the sky rms.
Postage stamp images were created by cropping a square region centered on the object with the side five times the semimajor axis of the SExtractor
estimate. We used the segmentation map to check if any pixels belonging to other objects were present in the square. If so, we filled these
pixels with the randomized background whose statistics were tuned to match the SExtractor estimate. 
Any objects whose half-light radii were less than the upper limit of the HST PSF value were discarded.

The next procedure is to plant these postage stamp UDF galaxy images onto each 4k $\times$ 4k CCD image. We looped over the list of the postage stamp images sequentially but
randomizing the location and the orientation. The loop ended when the desired number density was reached. A large fraction of the pixels were
the result of the superposition of many different postage stamp images. Thus, it is apparent that the noise level of these pixels vary slightly depending
on the number of overlaps. However, because we later introduce much larger additional noise to simulate the 15 s exposure data,
this noise variation is negligible.

A stellar count study based on the SDSS catalog (Juri{\'c} et al. 2008) shows that
a single 4k $\times$ 4k CCD image ($\sim14\arcsec\times14\arcsec$) in 15 s exposure at the Galactic north pole
is expected to contain at least $\sim140$ high S/N stars ($\geq 30\sigma$), 
which can be used to model the PSF. However, in the current simulation 
we conservatively distributed 40 point sources per CCD in such a way that 
the S/N of those stars remain above 30 in 15 s exposure, assuming the following differential star count: $\phi(m_r) \propto 10^{0.2 m_r}$;
we find that PSF reconstruction with different magnitude distribution does not impact the quality of the PSF reconstruction 
as long as all the stars used have S/N$>30$.

\begin{figure}
\includegraphics[width=18cm]{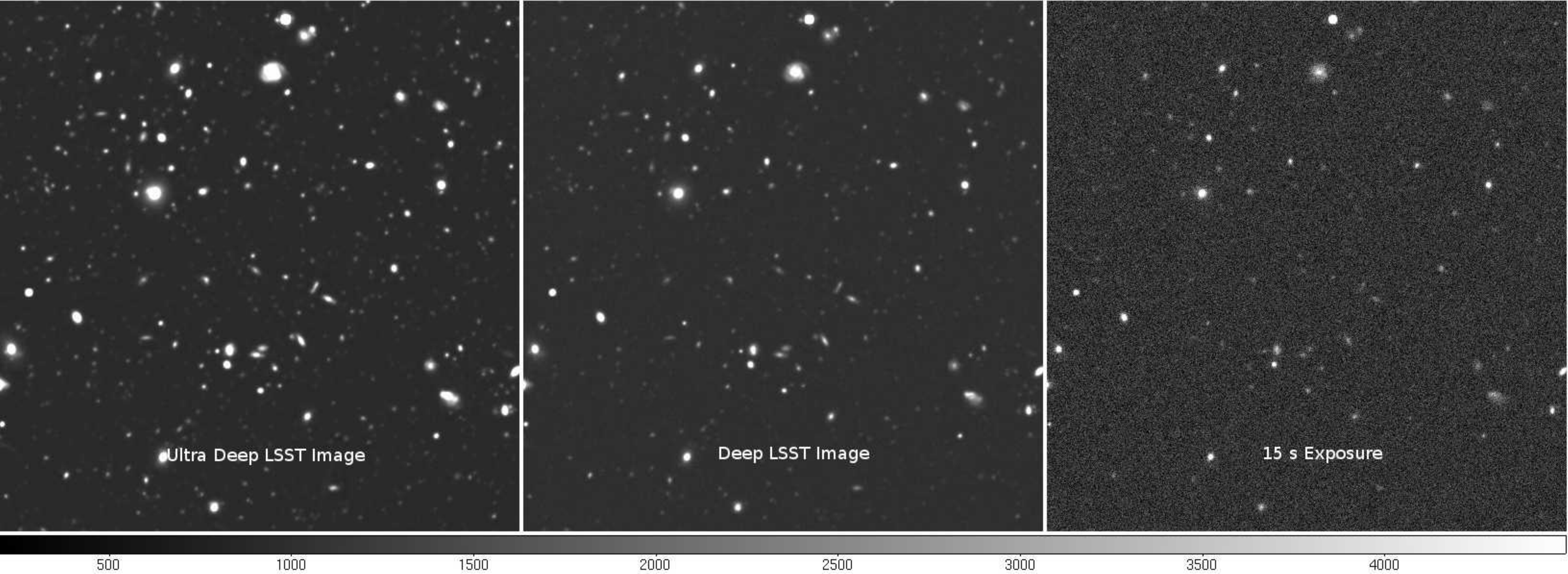}
\caption{Simulated LSST image. We display only a $\sim100 \arcsec \times 100 \arcsec$ cutout to show the detail.
The left panel image is the result prior to noise addition, about 2 mag deeper than the median depth (6000 s, $\sim27.5$ at 5 $\sigma$)
shown in the middle panel.
The nominal 15 second exposure case is shown in the right panel.
\label{fig_simulated_lsst}}
\end{figure}

Prior to the noise generation, the ``perfect'' 4k $\times$ 4k image must be convolved with the position-dependent PSF of the CCD.
Because our PSF is represented by 20 coefficients and 20 basis functions, it provides us with a convenient way to implement the spatially varying convolution.
To begin with, we convolve a 4k$\times$4k seeing-free astronomical image with 20 eigenPSFs via FFT. The results of this operation are 20 4k$\times$4k images.
Then, using the resulting 4th order polynomial fitting we generate another 20 4k$\times$4k images with each pixel containing 
the coefficient (amplitude) of the eigenPSF. Next, we multiply each eigenPSF-convolved image with
the matching coefficient image, obtaining 20 new 4k$\times$ 4k images. 
These 20 images are simply stacked together to produce one 4k$\times$4k image. This image is not yet the final
image that we desire to generate because the 20 eigenPSFs are derived from the mean-subtracted
PSFs. Therefore, we need to convolve the seeing-free 4k$\times$4k image with the mean PSF, and add the result
to the former 4k$\times$4k image to yield the final 4k$\times$4k LSST image.
Figure~\ref{fig_simulated_lsst} displays a $100\arcsec\times100\arcsec$ cutout from one of the 189 4k $\times$ 4k images.
The noise-free (except for the existing low-level noise in the ACS UDF image) LSST-seeing-convolved image is shown
in the left panel. This ``ultra deep'' image is about 2 mag deeper than the
median depth LSST image (middle) after the completion of the nominal 10 year mission with 200 exposures per sky patch.
The nominal 15 s LSST single exposure simulation is shown in the right panel.

\section{SIMULATED IMAGE ANALYSIS AND RESULTS}
\label{section_analysis}
\subsection{PSF Reconstruction}
\label{section_psf_reconstruction}
The procedure to apply PCA to the simulated LSST image containing galaxies and stars is similar to the one discussed in the case of the input PSF modeling.
The difference is that now we have far fewer stars, realistic noise, and neighboring objects. As a
matter of course, this will provide the reconstructed PSF with fewer degrees of freedom in the description of the
spatial variation. Hence, if the intrinsic PSF variation within a CCD is more complicated than what the reconstructed
model can predict, the weak-lensing science with LSST will be substantially compromised. In this section,
we present the detailed comparison between the reconstructed PSF and the input PSF in order to address this
critical issue.

\begin{figure}
\includegraphics[width=18cm]{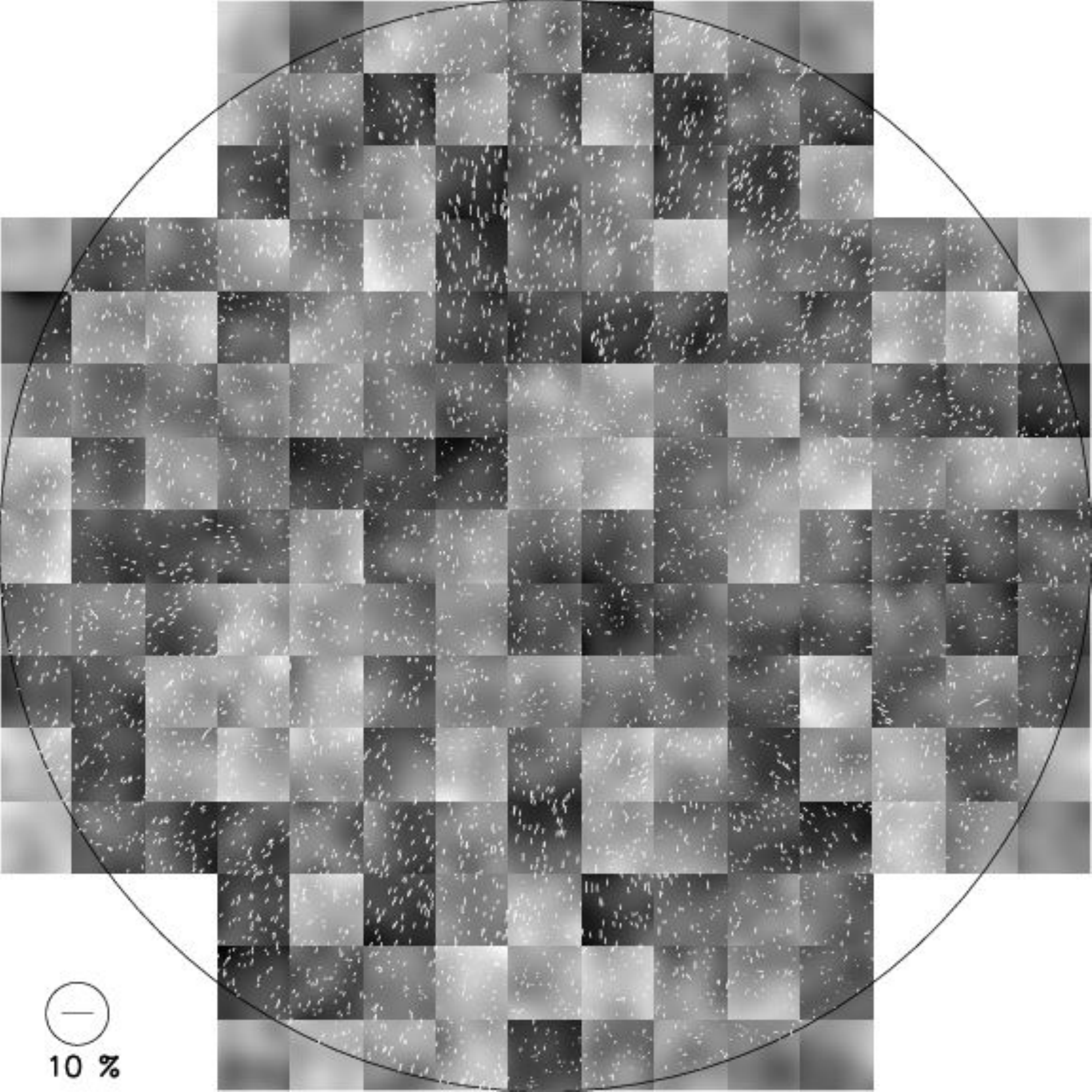}
\caption{Ellipticity distribution in simulated LSST images. We display ``whiskers'' overlaid on the LSST focal plane that we use for the input PSF generation.
Apart from the magnitude, the PSF pattern is similar to the one in the case when only optical aberration and focal plane height errors are considered 
(Figure~\ref{fig_ellipticity_focal_plane_no_atmos}). The reduction in the magnitude demonstrates that the atmospheric turbulence mostly
circularizes the PSF rather than introduces additional anisotropy. Because we simulate the sky near the zenith, the ellipticity
due to the atmospheric dispersion is very small in the current case.
Note that the correlation of the ellipticity change with the focal plane height variation is still visible.
\label{fig_ellipticity_focal_plane}}
\end{figure}

Astronomical objects in the simulated 189 4k$\times$4k images were detected with SExtractor. Stars were selected
by utilizing half light radius versus magnitude relation as performed in typical lensing analysis. The stellar
locus is well defined for bright stars. In addition, using
ellipticity, comparison of half-light radius with FWHM measurement, and source extraction history (e.g., presence
of close neighbor, edge clipping, etc.), the contamination becomes negligible in the current 10 sq. degree simulation. 
The cutout size is chosen
to be $31\mbox{pixel}\times31\mbox{pixel}$ ($6.2\arcsec\times 6.2\arcsec$) after the star is sub-pixel shifted
to place the centroid on a center of pixel. The ellipticity of the selected stars are displayed in Figure~\ref{fig_ellipticity_focal_plane}.
We overlay the ``whiskers'' over the focal plane image that we use for the input PSF generation.
The pattern of the PSF variation resembles the case displayed in Figure~\ref{fig_ellipticity_focal_plane_no_atmos}, where no
atmospheric effect is included. However, the reduction of the magnitude is apparent due to the circularization
effect of the atmospheric turbulence. We observe that no significant anisotropy is introduced by the moving phase screens.
The PSF elongation in the zenith angle direction is very small here because we choose a telescope
pointing to be close to the zenith.

\begin{figure}
\includegraphics[width=18cm]{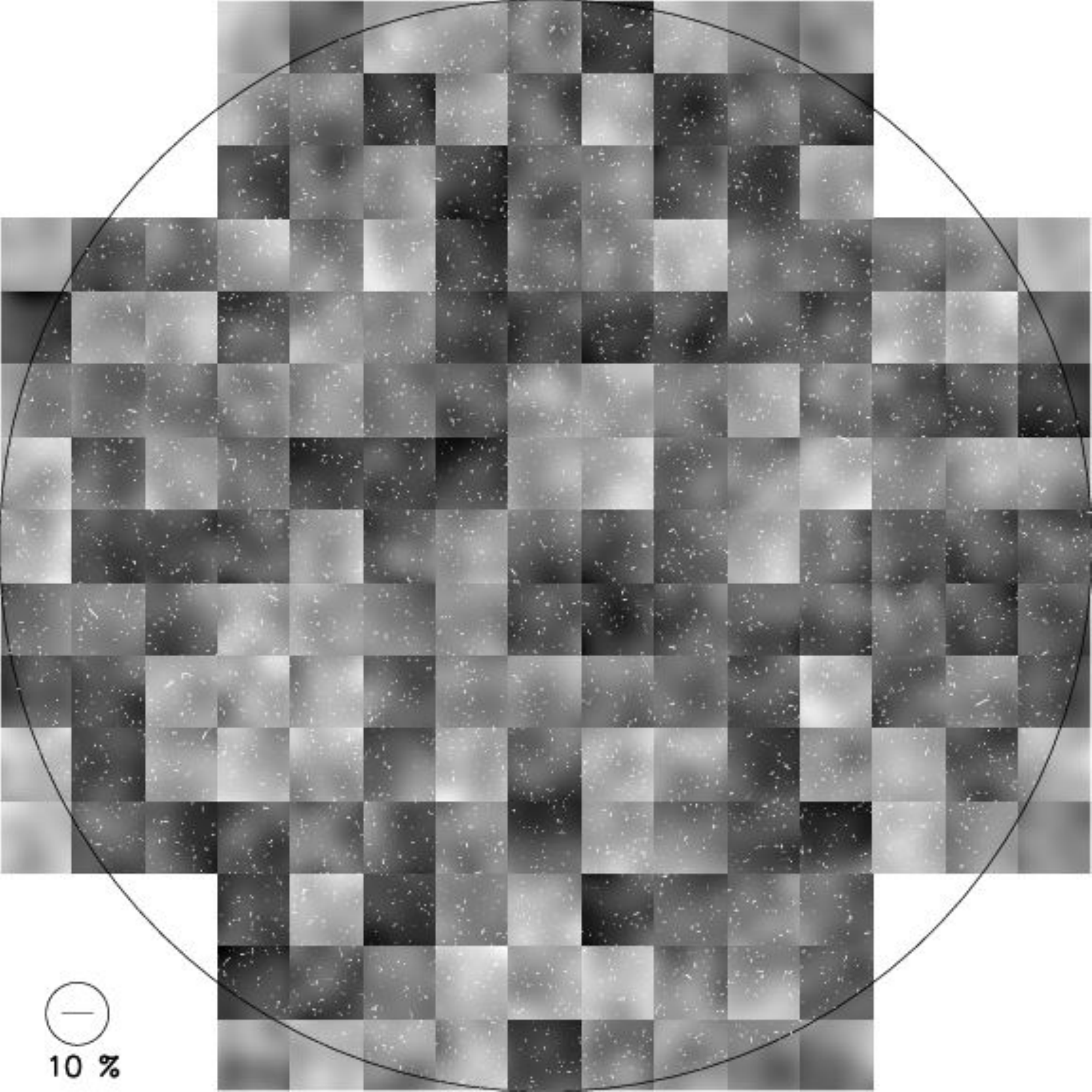}
\caption{Residual Ellipticity distribution. We subtracted the model ellipticity from the observed (simulated) ellipticity. Hence,
the length of the whiskers shows the magnitude of the residual ellipticity $(\delta e_{+},\delta e_{\times})$. 
\label{fig_residual_ellipticity_focal_plane}}
\end{figure}

\begin{figure}
\includegraphics[width=8cm]{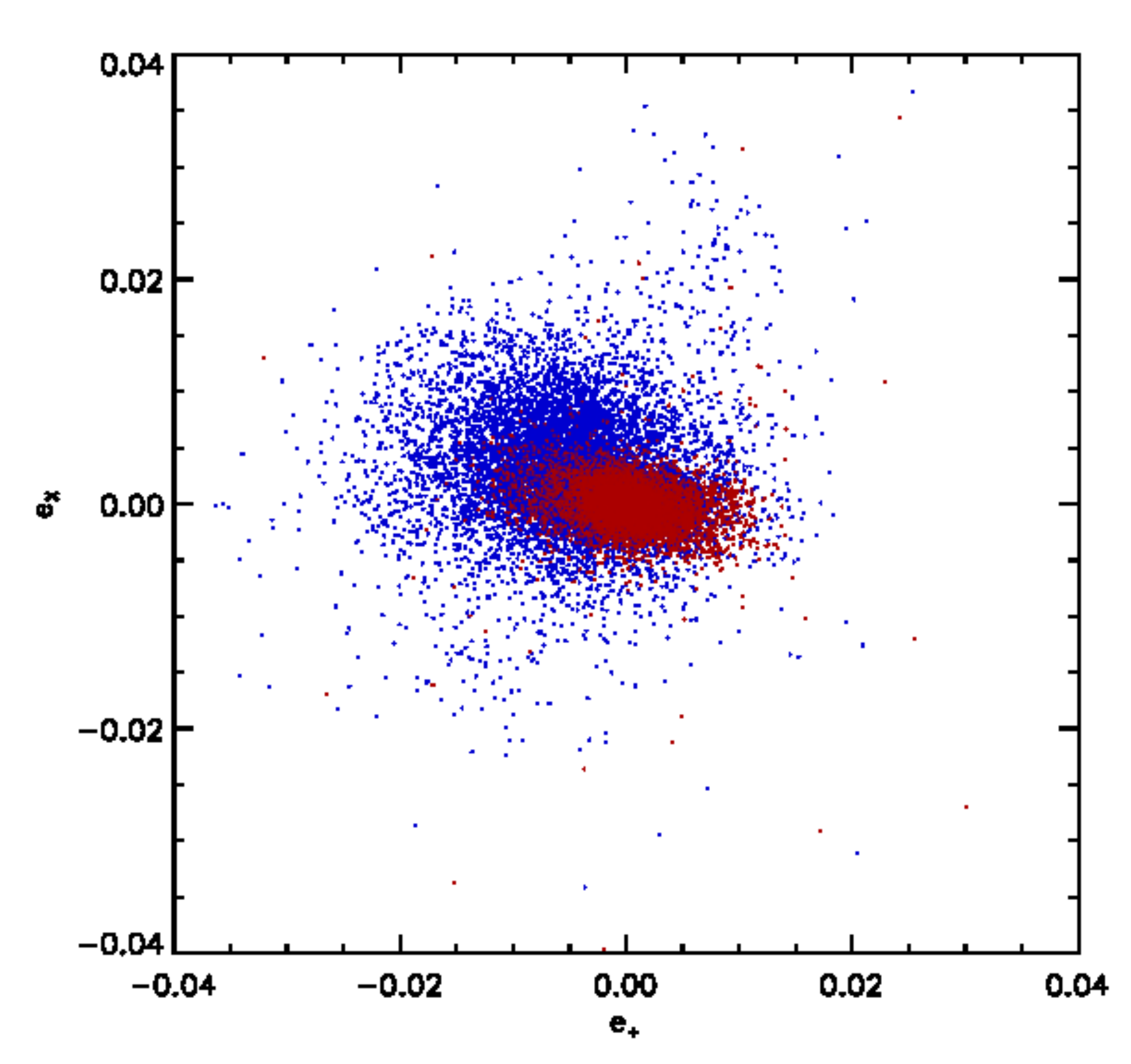}
\caption{Ellipticity components of PSF. The blue and red dots represent the ellipticity before and after the PSF correction, respectively.
\label{fig_ellipticity_component}}
\end{figure}

We derive eigenPSFs for each CCD separately and use 3rd order polynomials to describe the spatial variation.
For the visualization of how well the reconstructed model represents the observed pattern, we display the residual
ellipticity distribution across the focal plane in Figure~\ref{fig_residual_ellipticity_focal_plane}. 
The residual in Figure~\ref{fig_residual_ellipticity_focal_plane} is the difference in ellipticity ($e_{+}$,$e_{\times}$) of the observed (simulated) stars and
the model PSF evaluated at the location of the stars. Consequently, any correct model should not only reduce
the size of the ``whiskers'' in Figure~\ref{fig_ellipticity_focal_plane}, 
but also make the correlation of their size and the orientation decrease significantly. Visual comparison
of Figure~\ref{fig_residual_ellipticity_focal_plane} with Figure~\ref{fig_ellipticity_focal_plane}
shows qualitatively that our interpolated PSF at the location of the stars closely mimics the 
pattern of the observed PSFs. One method to quantitatively examine the reduction of the whiskers
is to compare the ellipticity component distribution in the two cases. 
In Figure~\ref{fig_ellipticity_component} the blue
dots represent the ellipticity components of the observed stars whereas the red dots indicate
those of the residuals. Note that the residual ellipticity distribution is much more compact
and is precisely centered on the origin $(e_{+},e_{\times})=(0,0)$. 

\begin{figure}
\includegraphics[width=8cm]{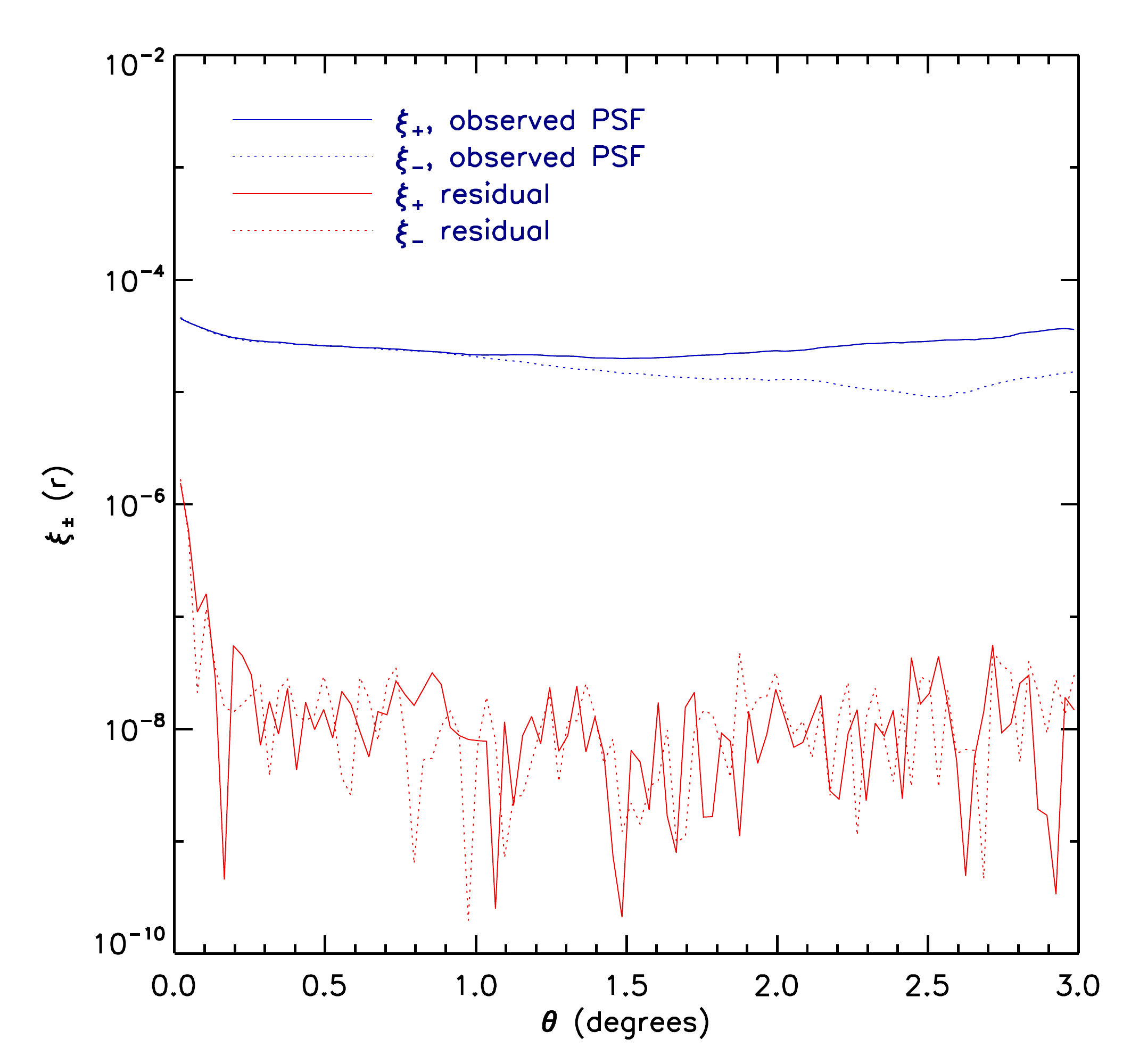}
\caption{PSF ellipticity spatial correlation. We examine the correlation function $\xi_{+}(r) $ for
PSF ellipticity. First, PSFs are paired, and then the tangential components with respect to the line connecting
the two PSFs are evaluated.  The amplitude of $\xi_{-}(r) $ is similar to that of $\xi_{+}(r)$, and is omitted for
clarity. Note the dramatic reduction of the spatial correlation for residual ellipticity of the stars.
\label{fig_ellipticity_correlation}}
\end{figure}

Figure~\ref{fig_ellipticity_correlation} displays the ellipticity correlation.
The spatial correlation of the observed ellipticity (before PCA correction) has an amplitude of $3\times10^{-5}$. 
For a sanity check, we compared this amplitude with the values that we obtain from real observations, and found that
they agree well. For example, the amplitude of the intrinsic PSF correlation before correction
is $\sim4\times10^{-5}$ and $\sim3\times10^{-5}$ in the Deep Lens Survey (DLS; Wittman et al. 2002) field from 
the 4m Kitt Peak telescope and 8m Subaru telescope images, respectively;  one should use caution when comparing these values to other published results because of different ellipticity definitions and weight functions.

Also displayed in Figure~\ref{fig_ellipticity_correlation}  is the ellipticity correlation function of the purely atmospheric PSF (Figure~\ref{fig_at_psf}), 
which is generated by ``turning off" optical aberrations. These PSFs are free from interpolation errors  and thus help us to assess
the amount of the atmospheric contribution to the PSF anisotropy. In particular, the comparison enables us to verify that
$\sim800$ stars per CCD is sufficient to model the PSF pattern within a CCD. If the atmospheric PSF possessed a significant
power on a small scale $r<1\arcmin$, those $\sim800$ stars could not adequately sample the PSF variation. 

Obviously, the amplitude $3\times10^{-5}$ of the ellipticity correlation for the observed PSF (blue) is
higher than the expected cosmic shear signal for the $\Lambda$ CDM universe at large angles.
This implies that without PSF
correction the galaxy ellipticity correlation will be dominated by the PSF systematics rather than by the
intrinsic lensing signal. It is reassuring that the correlation function of the residual ellipticity
components has an amplitude of $\sim10^{-8}$, which is more than 3 orders of magnitude reduction in correlation and
also more than 2 orders of magnitude below the anticipated cosmic shear lensing signal (see Rowe 2010 for
the discussion and use of the residual ellipticity correlation in detail). We find similar improvement when we apply the PCA
PSF corrections to the DLS data.

It is important to remember that the above PSF correlation is measured on a single 15 s exposure image.
In fact, LSST will visit the same patch of the sky roughly 200 times in each filter used for weak-lensing with two 15 s exposures.
Before each exposure the telescope optics is updated from the wavefront sensing system.
The pupil rotation, sky rotation, and $x-y$ dithers will mix different systematic PSF patterns and reduce the
intrinsic ellipticity of the PSFs in the final stack. In addition, this will also 
remove the spatial correlation that we saw within a single CCD even before we apply PSF corrections.

\begin{figure}
\includegraphics[width=18cm]{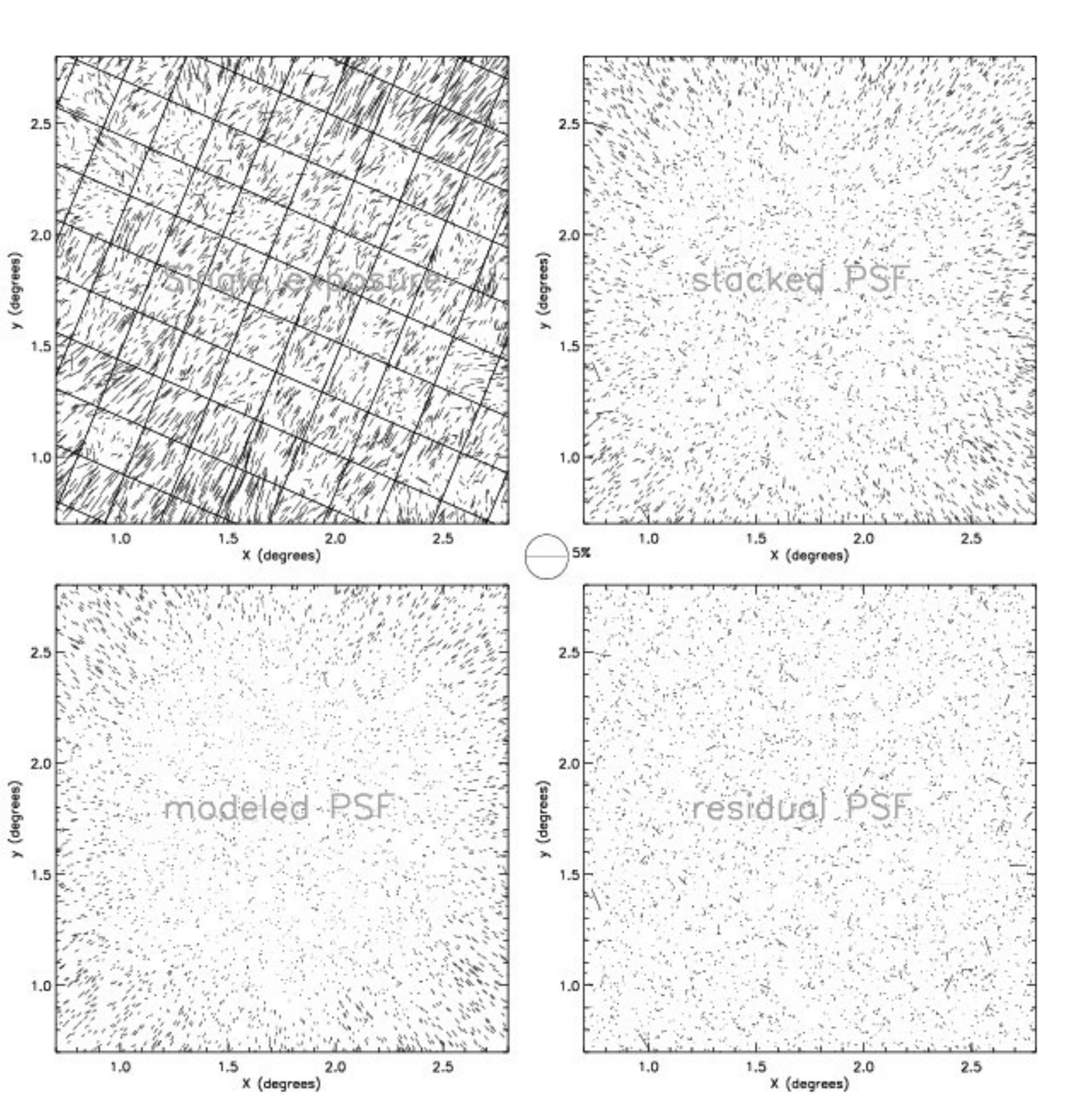}
\caption{Circularization of LSST PSF from field rotation. 
We illustrate a toy-model observing scenario, where a same patch of the sky is
visited 100 times with randomized roll angles. The focal plane height fluctuation is assumed to remain
constant during the entire mission whereas the atmospheric turbulence changes with time.  Post active optics residuals are
small by comparison. We display only the fraction of the focal plane  
for easy viewing.
The observed PSF ellipticity
(upper right) is much smaller than seen in individual single 15 s observations (upper left). The lower left panel shows the reconstructed PSF by first performing
PCA on each exposure and then stacking all model PSFs with correct application of field rotation. The residual PSF ellipticity
is displayed in the lower right panel.
\label{fig_field_rotation}}
\end{figure}

Although one can come up with an optimized operation scheme, which takes full
advantage of hundreds of visits, we observe that even a simple random field rotation
can dramatically reduce the intrinsic PSF ellipticity correlation. Figure~\ref{fig_field_rotation}
schematically displays this simple observing pattern and the resulting PSF ellipticity distribution.
We illustrate an over-simplified observing scenario, where a same patch of the sky is
visited 100 times with randomized roll angles. 

In order to create the stack image out of these 100 frames,
we arbitrarily choose one frame as a reference and rotate the rest of the frames for the alignment.
Stacking rotated images are highly non-trivial in the presence of anisotropy in pixel boundaries, atmospheric dispersion, 
satellite trails, cosmic rays, detector response variation, internally scattered light/ghosts, etc\footnote{While beyond the scope of this paper,
our ``Multi-Fit'' approach (Tyson et al. 2008) to galaxy photometry and shape measurement fits a galaxy morphological model to all
the exposures, and thus can avoid the propagation of these issues into the LSST weak-lensing science.}
Numerous schemes exist for fractional pixel handing. We use
bi-cubic interpolation (Press et al. 1992), which closely mimics the ideal sinc interpolation or
its variations (e.g., Lanczos kernels), for both stacking images and constructing PSFs;
the merits of using Lanczos kernels are discussed in the literature (e.g., Jee et al. 2007; Bertin 2010).
The atmospheric dispersion is very small in the current simulation because we simulated pointings near the zenith. 
Thus, the dispersion correction is omitted in the current simulation. However, we think this atmospheric
dispersion and other aforementioned real world problems must be handled with care because any small bias may
appear as second-order systematics after analysing many billion galaxies in the 40,000 sq. degree area.
These issues will be discussed in our future publications.

We use the single realization of the focal plane (shown in Figure~\ref{fig_focal_plane}) in the above multi-epoch simulation although we randomize the phase of the atmospheric turbulence screen for each epoch.
It is important to realize that this simulation setup for the telescope and camera closely reflects the planned operation scheme of LSST. The LSST camera will
maintain its fixed focal plane configuration throughout the mission, and the mirror deformation and alignment changes will be
corrected with actuators to $\lesssim0.2\mu$m accuracy (Manuel et al. 2010). That is, the difference in aberration for different visits is small
compared to the aberration introduced by the focal plane non-flatness.
In principle, this small wavefront reconstruction error may slightly alter the resulting telescope aberration patterns exposure by exposure.
However, the wavefront reconstruction error in different exposures are random, and stacking many exposures will reduce the correlation, circularizing the final PSFs to our advantage.

\begin{figure}
\includegraphics[width=8cm]{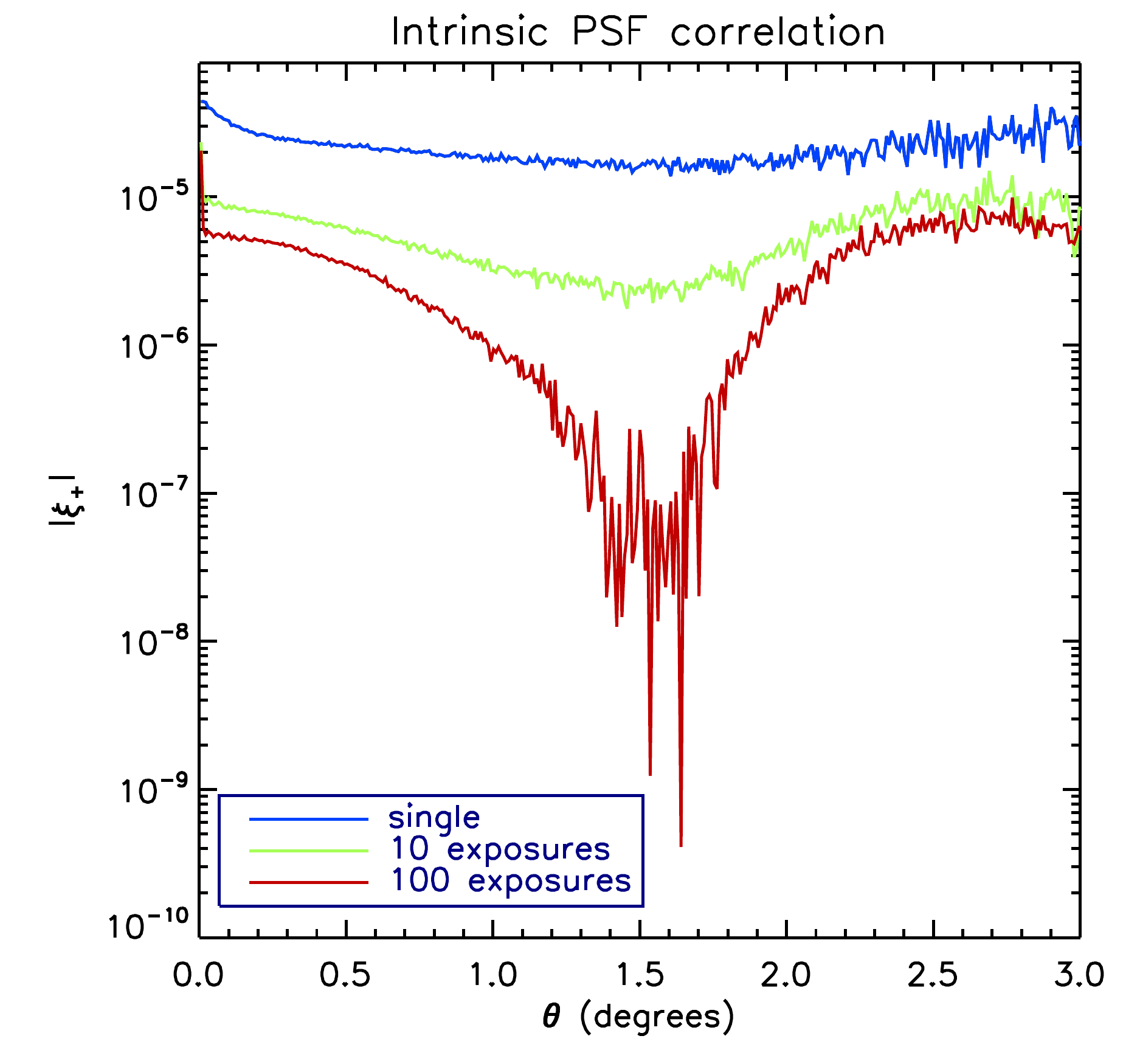}
\includegraphics[width=8cm]{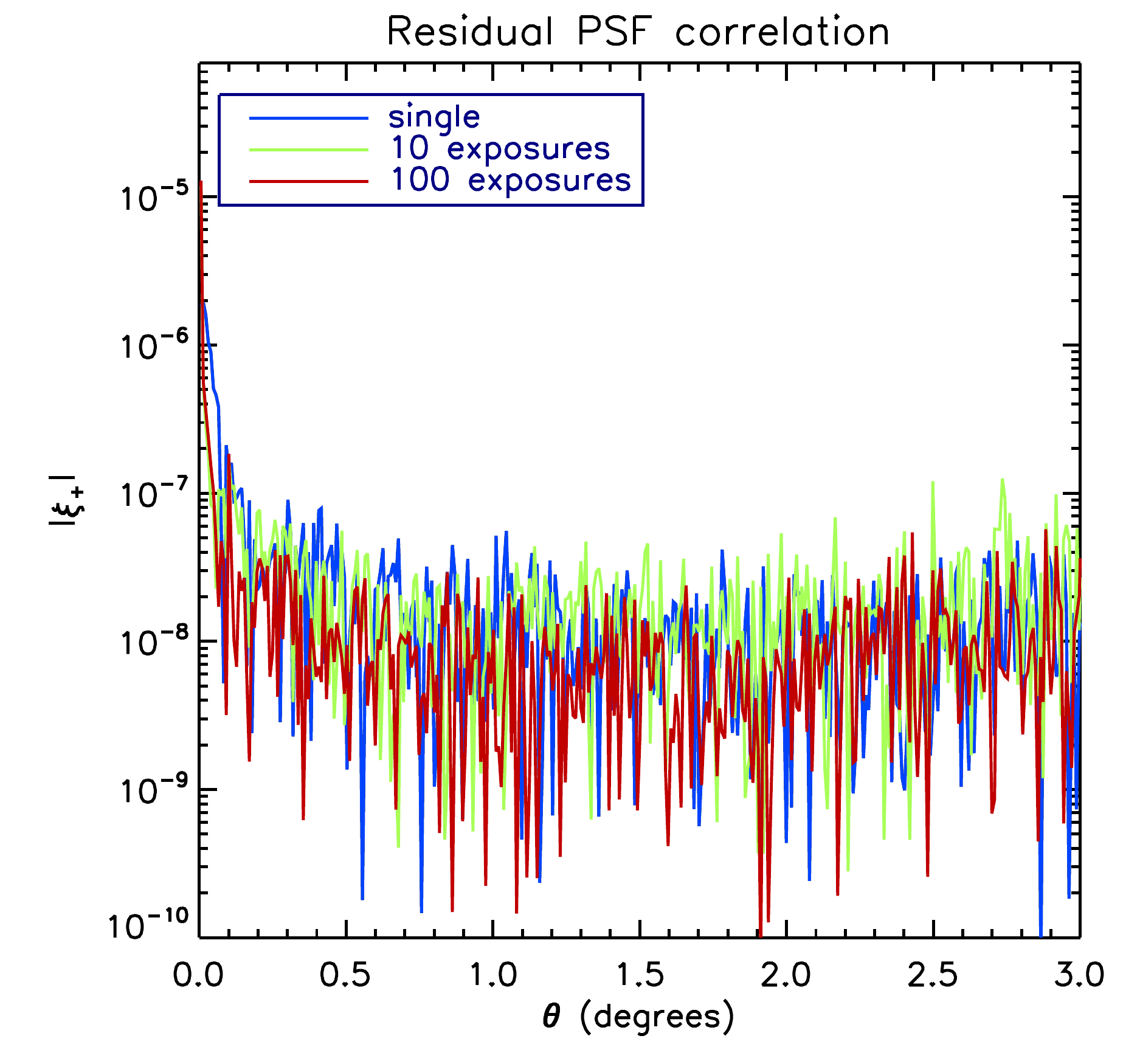}
\caption{Same as Figure~\ref{fig_ellipticity_correlation} but for the case that a same patch of the sky is
visited multiple
times with randomized roll angles. Stacking different PSF patterns reduces the correlation
in intrinsic PSF (left). The residual PSF correlation function (right) remains at the similar level as in Figure~\ref{fig_ellipticity_correlation}, which
is not surprising because the amplitude is already at the statistical limit.
\label{fig_ellipticity_correlation_rotation}}
\end{figure}

In Figure \ref{fig_field_rotation} the observed PSF ellipticity after stacking
(upper right) is much smaller than what is seen in individual single 15 s observations (upper left). 
The lower left panel shows the reconstructed PSF by first performing
PCA on each exposure and then stacking all model PSFs after correct application of field rotation;
the rotated PSFs are created by re-sampling with bi-cubic interpolation, which is identical
to the procedure in the image stacking above.
Stacking different PSF patterns reduces the correlation
in observed PSF (Figure~\ref{fig_ellipticity_correlation_rotation}.)  However, it does not bring down the amplitude
below $10^{-7}$ over the entire range of angular scales. This is because our toy model observing scheme without $x-y$ dithers does not
mix different PSFs efficiently.

\subsection{Null Test with Galaxies}

The high-fidelity reconstruction of the observed PSF through PCA demonstrated in \textsection\ref{section_psf_reconstruction}
hints at the prospect that the resulting galaxy shear measurement can be carried out with similarly small systematic errors.
However, there remain several issues. First, anisotropic PSFs
change the pixel noise properties also anisotropically, and thus even with the knowledge of the perfect PSF
the resulting galaxy shape measurement can be still biased along the elongation of the PSFs. Related to this point is
the centroid bias (e.g.,Kaiser 2000; Bernstein \& Jarvis 2002), which causes 
the object centroid to be more uncertain in the direction of the PSF elongation.
Second, our comparison between the model and the observation is limited to the PSFs at the location of the stars. If the
intrinsic spatial variation of the PSF has higher frequency than the order of the polynomial (3rd order here), it is
possible that galaxy shape measurements would be carried out with suboptimal PSFs.
Third, apart from PSF modeling, the difference in galaxy radial profile between assumption and reality can
cause biases in shear extraction as suggested by Bernstein (2010). Fourth, it is not yet clear how to handle
the pixelization effect optimally. Because photons are collected in finite CCD pixels, 
any shape measurement algorithm needs to consider this procedure to avoid bias. Similar to the convolution
by seeing, the pixelization also somewhat destroys the information, especially at the core of object, where
the radial profile is most steep. [However, the low surface brightness limit of each exposure allows shape measurement at
large radius.]  Fifth, object detection from real astronomical images involves many subtleties such
as deblending, spurious object identification, etc. The current algorithms need to be
improved substantially to minimize the contamination from these undesired detections or at least to quantify
the fraction more precisely in order to estimate the dilution of the lensing signal.

In terms of shear extraction, the above concerns can be largely divided into two issues: the removal of the anisotropic bias and the calibration
bias. For convenience, many authors (Heymans et al. 2006; Massey et al. 2007; Bridle et al. 2010)
formulate the two issues with the following expression:
\begin{equation}
\gamma^o - \gamma^t = m \gamma^t + c,
\end{equation}
\noindent
where $m$ and $c$ are the calibration and anisotropic biases, respectively, and $\gamma^o$ and $gamma^t$ are
the observed and true shears, respectively.
Despite the difficulties mentioned above, as a last resort the calibration bias or multiplicative factor $m$  can be
determined through extensive image simulations. We prefer a method which
provides self-contained calibration factor and relies on the image simulation result only as a confirmation tool.
Within the scope of this paper, we address the second issue, i.e., the removal of anisotropic bias, by performing ``null'' test
on the simulated LSST images. In particular, we focus on how well a simple field rotation can remove this additive factor.

For the measurement of the galaxy ellipticity, we fit PSF-convolved elliptical Gaussians (e.g., Jee et al. 2009) to galaxies. As mentioned
in \textsection\ref{section_psf_reconstruction}, the PSF on the final stack is constructed from the
superposition of the 100 PSFs from all visits. We discard objects whose half-light radius is less than
1.2 times the median value of the stars or ellipticity measurement errors greater than 0.2.
The resulting number of usable galaxies is $\sim40$ per sq. arcmin for the depth at the completion of the nominal LSST survey, which
meets the weak-lensing science requirement (LSST Science Collaborations et al. 2009). The exact value of the selection
criteria should be determined in conjunction with the study of shear calibration, and thus will be
the subject of our future publications.

As no shear is introduced, the galaxy correlation function should possess amplitudes consistent with the shot noise.
One should remember that since the intrinsic ellipticity of individual galaxies are at least a factor of 10 larger than that of PSFs, 
 the correlation function within a single LSST field will not approach the level of the residual PSF correlation function.

\begin{figure}
\includegraphics[width=8cm]{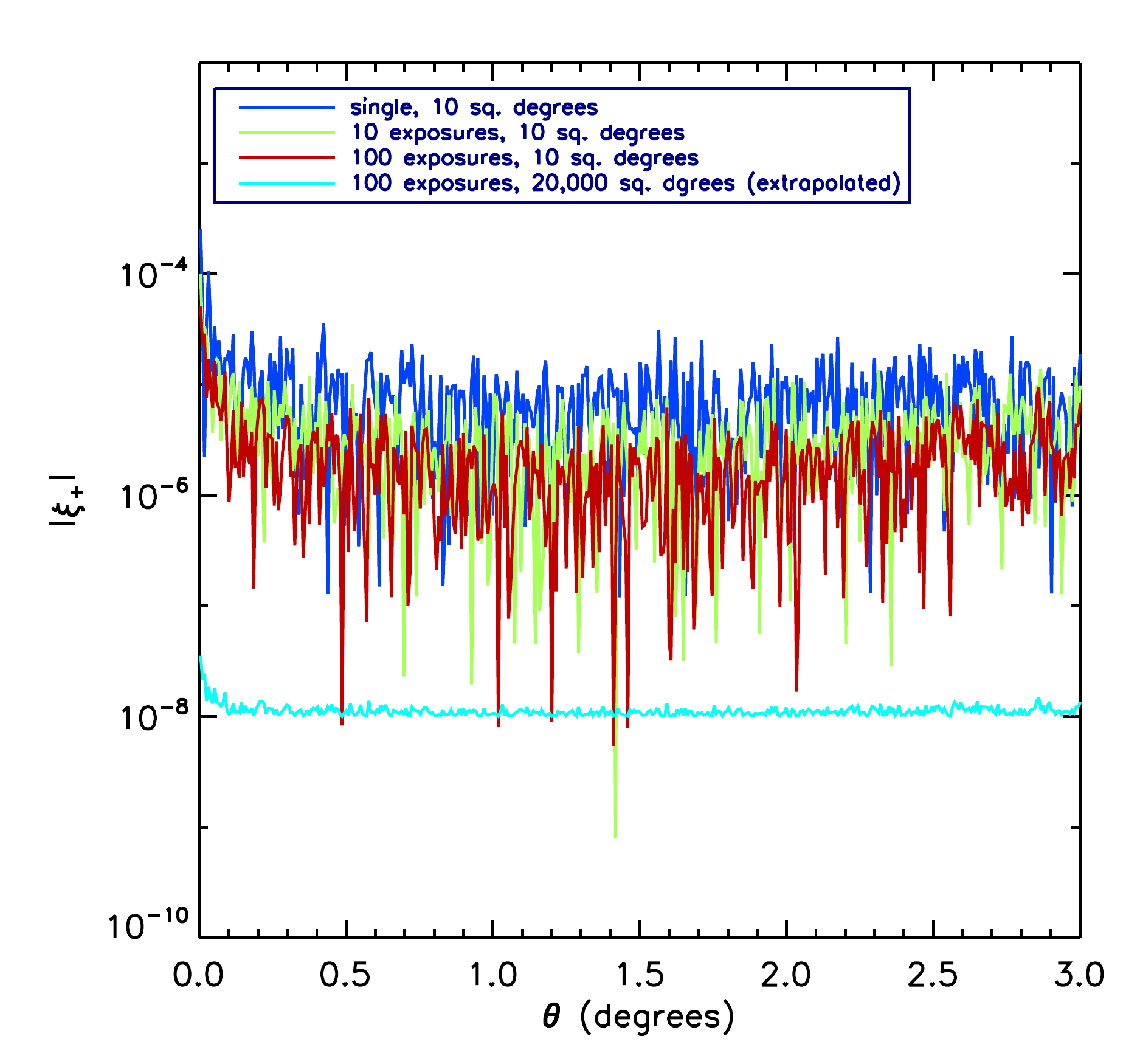}
\caption{Null test with galaxy ellipticity correlation for a single field of view and for
20,000 sq. degrees. If anisotropic PSF effects are corrected, the correlation
amplitude should simply reflect the statistical noise as observed.
As the number of exposures increases,
the correlation amplitude goes down because the residual PSFs become more uncorrelated (systematic noise) and also
the increased depth gives higher number of galaxies usable for shear measurements (statistical noise).
The number density of usable galaxies for weak-lensing shear extraction in one, 10, and 100 exposures are 10, 24, and 42 galaxies per sq. arcmin,
respectively.
The area of the simulated image is $\sim10$ sq. degrees. If we extrapolate the current result to the case, where
all 20,000 sq. degrees of the area are taken, the expected amplitude of the residual galaxy shear correlation is below $\sim10^{-8}$ (cyan).
\label{fig_g_correlation}}
\end{figure}

\begin{figure}
\includegraphics[width=8cm]{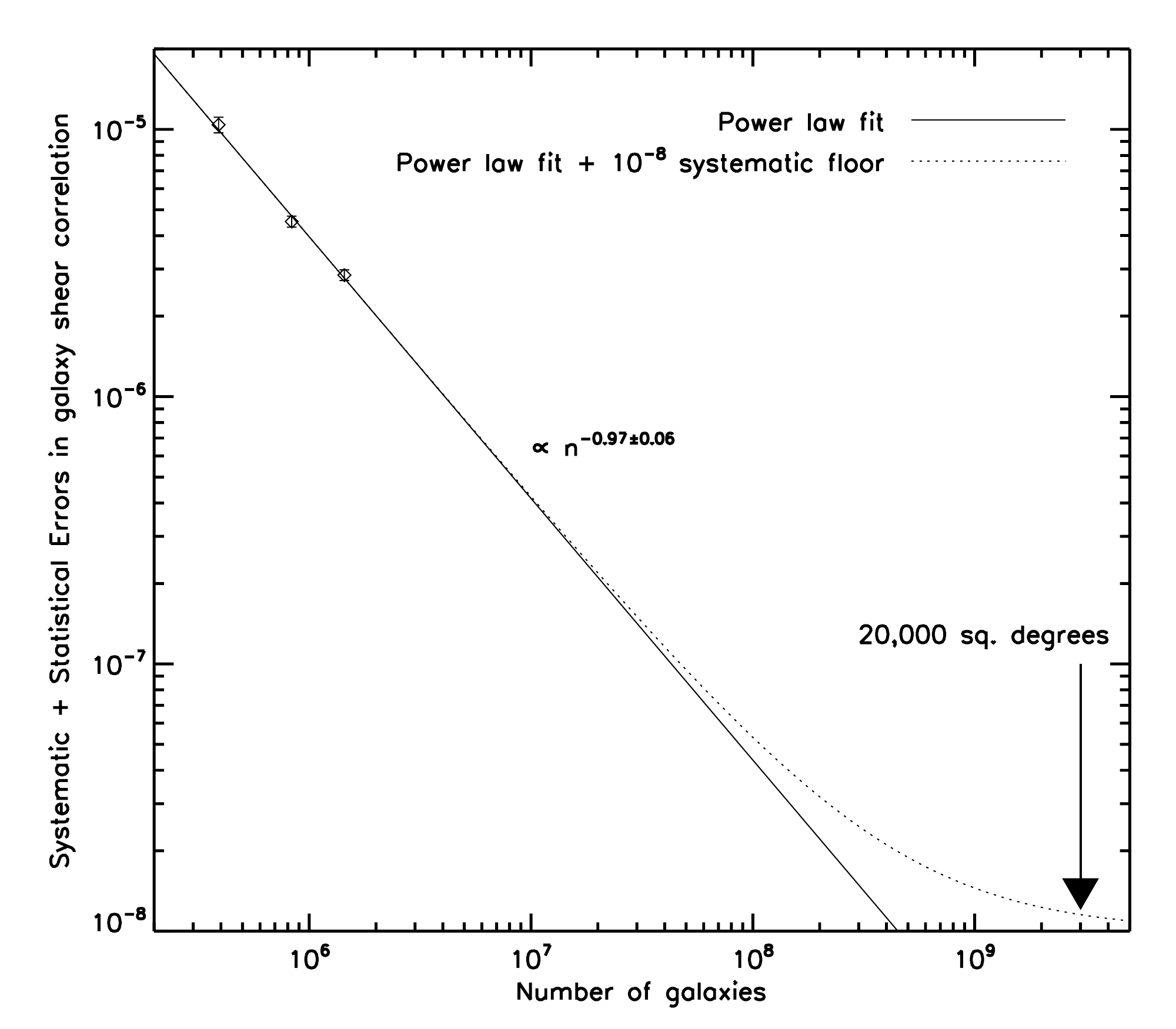}
\caption{Total (systematic + statistical) errors in galaxy shear correlation. The power-law fit 
gives a slope $-0.97\pm0.06$, consistent with the $\propto 1/n$ behavior of the statistical noise (solid line).
A systematic noise, although not detected in the current 10 sq. degree simulation, will 
determine the minimum noise level when the nominal 20,000 sq. degree survey is completed ($\sim3$ billion galaxies).
The dotted line shows the modification when we assume a $10^{-8}$ systematic noise floor. The expected
cosmological signal is greater than $10^{-6}$ over these angular scales.
\label{fig_shear_correlation_error}}
\end{figure}

In Figure~\ref{fig_g_correlation}, we present this ``null'' test result with galaxy ellipticity for the 10 sq. degree field.
If anisotropic PSF effects are corrected, the correlation amplitude should simply reflect the statistical galaxy shot noise as observed.
As the number of exposures increases,
the correlation amplitude goes down because the residual PSFs become more uncorrelated (systematic noise) and also
the increased depth gives higher number of galaxies usable for shear measurements (statistical shot noise).
The number density of galaxies with sufficiently high S/N for ellipticity measurement for a single, 
10, and 100 exposures are 10, 24, and 42 galaxies per sq. arcmin,
respectively. We find that the mean amplitudes of the galaxy ellipticity correlation functions
are closely proportional to the inverse of these number densities (Figure~\ref{fig_shear_correlation_error}).
This tight relation implies that the level of the systematic error for the 100 exposure data set
is much less than that of the statistical error ($\lesssim10^{-7}$). 
The expected signal for a $\Lambda$CDM cosmology corresponds to a galaxy ellipticity correlation $>10^{-6}$ over
this range of angular scale.
The exact level of the systematic floor can be estimated only by extending
the area of the simulation by a large amount. For the case that the amplitude
of the PSF ellipticity correlation function $\sim10^{-8}$ in Figure~\ref{fig_ellipticity_correlation_rotation} happens to be the
fundamental limit, we estimate the extrapolated correlation function for the 20,000 sq. degree area survey (cyan line in Figure~\ref{fig_g_correlation}).

\section{CONCLUSIONS AND SUMMARY}
\label{section_conclusion}
We have investigated how the atmospheric turbulence and the telescope aberrations will impact the PSF of the active optics LSST in the context
of weak lensing science. The atmospheric turbulence simulated by six moving Kolmogorov phase screens mainly contributes
to the broadening of the diffraction-limited PSF (that would have been obtained in the absence of the atmosphere) without
significantly inducing additional anisotropy. The FWHM $\sim0.7\arcsec$ of the simulated PSF nicely agrees with the mean value determined by the
LSST seeing monitoring program at Cerro Pach\'{o}n. 
The instrumental aberrations place potential challenges to precision weak-lensing analysis with LSST because of the fast f-ratio of the optics.
Our simulation shows that the resulting shallow focal depth creates abrupt changes in PSF pattern across CCD borders.
As the $3.5\degr \times 3.5\degr$ focal plane will be tiled with 189 CCDs each having its own surface distortion (i.e., potato-chip effect), 
the PSF pattern on the entire focal plane is very complex.

However, we demonstrate that the complexity of the PSF pattern over the entire focal plane can be nicely described by
interpolation with basis functions derived by PCA. In order to account for the discontinuity across the CCD borders with
polynomials, the basis functions and the polynomial coefficients are determined separately for each CCD.
This PSF recovery test is carried out with simulated LSST images containing realistic galaxies, as well as stars in the
presence of photon noise and neighboring objects. The residual PSF ellipticity correlation function has
amplitudes of $10^{-8}$ over the $r=0.2\sim3.0\degr$ range, significantly lower than the amplitudes of the cosmic shear and
the raw PSF ellipticity correlation. 

It is critical to remember that in actual LSST operations the same patch of sky will be
visited multiple times with different combinations of pupil rotation, sky rotation, and $x-y$ dither. Hence, this will
substantially reduce our burden in modeling the PSF for individual exposures because the mix of
different PSF pattens will circularize the PSF in the stacked image and also decreases the spatial correlation.
To illustrate how much the stacking procedure can improve the systematics, we simulate
a toy case, where 100 visits are randomly rotated around the center of the field. The reduction in both ellipticity and
correlation prior to any PSF modeling and correction is dramatic even in this simplistic scenario.
The ``null'' test with galaxies in the stacked image indicates that the amplitude of the
galaxy shear correlation is consistent with the level of the shot noise, indicating that the systematic error
is not detectable within a single $\sim10$ sq. degree field with 100 exposures.

The PSF anisotropy correction (additive factor in shear recovery) is one of the most
poorly investigated effects, and the literature disagrees on the best approaches.
Our current simulation shows that the anisotropic PSF effects can be sufficiently corrected 
with the current design of LSST. This improvement is enabled by both a new PCA algorithm and the
large \'{e}tendue and active optics of LSST capable of turning systematics into statistics.
In a forthcoming paper, we will extend the current simulation with the addition of
input galaxy shear and discuss the accuracy of the shear recovery with existing algorithms. 

M. James Jee acknowledges support for the current research from the TABASGO foundation in the form of
the Large Synoptic Survey Telescope Cosmology Fellowship. We thank Lynne Jones for help with stellar
density estimation at the Galactic pole. We thank Chuck Claver for providing the ZEMAX model of the LSST.
We thank David Wittman, Andrew Connolly, Andrew Becker, and 
Andrew Rasmussen for useful discussions on focal plane non-flatness and PSF impact. We thank Perry Gee
for help in the parallelization of our simulation code.  Finally, we thank the anonymous referee who helped
us to improve the quality of the paper significantly.

\clearpage

\clearpage

\begin{deluxetable}{lcc}
\tabletypesize{\scriptsize}
\tablecaption{Focal plane error budget (peak-to-valley)}
\tablenum{1}
\tablehead{\colhead{Source} & \colhead{Specification (Maximum Values)} & \colhead{Values used for simulation} \\}
\tablewidth{0pt}
\startdata
Global focal plane adjustment & 10 $\mu$m & 10 $\mu$m \\
Raft height error\tablenotemark{1} & 4.8 $\mu$m & 6.5  $\mu$m\\
CCD-to-CCD height variation & 4 $\mu$m & 10 $\mu$m\\
CCD tilt & $2.5 \times 10^{-4}$ rad & $2.5 \times 10^{-4}$ rad\\
Potato chip effect\tablenotemark{2} & 4 $\mu$m & 5 $\mu$m \\
\enddata
\tablenotetext{1}{$3\times3$ CCDs will be attached to one raft.}
\tablenotetext{2}{1-2 cycles within a CCD}
\end{deluxetable}

\end{document}